\documentclass[12pt, english]{article}
\usepackage[T1]{fontenc}
\usepackage[english]{babel}
\usepackage[letterpaper]{geometry}

\newcommand{\mar}{1.0in}
\usepackage{fourier}
\usepackage{pdflscape}

\usepackage{microtype} 
\usepackage{graphicx}
\usepackage{array}
\usepackage[longnamesfirst,authoryear]{natbib}
\usepackage[textfont=sc,labelfont=bf]{caption}
\usepackage{setspace}
\usepackage{multirow}
\usepackage{amsthm}
\usepackage{amsmath}
    \allowdisplaybreaks[1] 
\usepackage{amsfonts}
\usepackage{amssymb}
\usepackage{svg}
\usepackage{pgf} 
\usepackage{bm}
\usepackage{paralist}
\usepackage{afterpage}
\usepackage{pdfpages}
\usepackage{xfrac}
\usepackage{arydshln}
\usepackage{tabularx}
\usepackage{epstopdf}
\usepackage{comment}
\usepackage{stackrel}
\usepackage[normalem]{ulem}
\usepackage{mathtools}
\usepackage{bbm}
\usepackage{booktabs}
\usepackage{subcaption}
\usepackage{float} 
\usepackage{array}
\usepackage{wasysym}
\usepackage{enumitem}
\usepackage{tikz} 
\usepackage{makecell} 
\usepackage{longtable}

\setlist[itemize]{leftmargin=*} 

\usepackage{titlesec}
    \titleformat*{\section}{\Large\bfseries}
    \titleformat*{\subsection}{\large\bfseries}
    \titleformat*{\subsubsection}{\large\bfseries}

\usepackage{xcolor}
    \definecolor{darkblue}{RGB}{30,30,130}
    \definecolor{darkred}{RGB}{130,10,10}
    \definecolor{midblue}{RGB}{120,120,255}
    \definecolor{midred}{RGB}{255,120,120}
    \definecolor{lightred}{RGB}{255,227,227}
    \definecolor{amaranth}{RGB}{230,43,79}
    \definecolor{cardinal}{RGB}{196,31,59}
    
    \definecolor{tab_blue}{RGB}{31,119,180}
    \definecolor{tab_orange}{RGB}{255,127,14}
    \definecolor{tab_green}{RGB}{44,160,44}
    \definecolor{tab_red}{RGB}{214,39,40}
    \definecolor{tab_purple}{RGB}{148,103,189}

\definecolor{matlab_1}{rgb}{0,0.447,0.741}
\definecolor{matlab_2}{rgb}{0.85,0.325,0.098}
\definecolor{matlab_3}{rgb}{0.929,0.694,0.125}
\usepackage[unicode=true]{hyperref}
\hypersetup{colorlinks=true,linkcolor=darkred,citecolor=darkred,urlcolor=darkblue}

\newcolumntype{L}[1]{>{\raggedright\let\newline\\\arraybackslash\hspace{0pt}}m{#1}}
\newcolumntype{C}[1]{>{\centering\let\newline\\\arraybackslash\hspace{0pt}}m{#1}}
\newcolumntype{R}[1]{>{\raggedleft\let\newline\\\arraybackslash\hspace{0pt}}m{#1}}
\DeclareMathAlphabet\mathbfcal{OMS}{cmsy}{b}{n}

\newcommand{\beq}{\begin{equation}}
\newcommand{\eeq}{\end{equation}}
\newcommand{\bea}{\begin{eqnarray}}
\newcommand{\eea}{\end{eqnarray}}
\newcommand{\ba}{\begin{array}}
\newcommand{\ea}{\end{array}}
\newcommand{\bit}{\begin{itemize}}
\newcommand{\eit}{\end{itemize}}
\newcommand{\ben}{\begin{enumerate}} 
\newcommand{\een}{\end{enumerate}}
\newcommand{\bpm}{\begin{pmatrix}}
\newcommand{\epm}{\end{pmatrix}}
\newcommand{\bbm}{\begin{bmatrix}}
\newcommand{\ebm}{\end{bmatrix}}

\makeatletter
\theoremstyle{plain}
\newtheorem*{prop*}{\protect\propositionname}
\makeatother
\providecommand{\propositionname}{Proposition}

\newif\ifreview
 \reviewfalse
\ifreview
\paperwidth=\dimexpr \paperwidth + 3cm\relax
\oddsidemargin=\dimexpr\oddsidemargin \relax
\evensidemargin=\dimexpr\evensidemargin + 3cm\relax
\marginparwidth=\dimexpr \marginparwidth + 3cm\relax
\usepackage{todonotes}
\else
\usepackage[disable]{todonotes}
\fi 

\usepackage{titling}
\title{\LARGE \textbf{Who's at Risk?  \\ Effects of Inflation on Unemployment Risk}\thanks{
\textbf{Disclaimer}: The views expressed in this paper are those of the authors and do not necessarily reflect the views and policies of the Board of Governors or the Federal Reserve System or the views of Citigroup and its affiliates.}}

\author{%
    Hie Joo Ahn\thanks{Federal Reserve Board of Governors, 20th Street and Constitution Avenue NW, Washington, DC 20551, U.S.A. Email: \protect\href{mailto:hiejoo.ahn@frb.gov}{hiejoo.ahn@frb.gov}} \\ [-0.25cm] 
	\normalsize Federal Reserve Board \and 
    Lam Nguyen \thanks{Citigroup, 6400 Las Colinas Blvd, Irving, TX 75039, U.S.A. Email: \protect\href{mailto: hoanglam296@gmail.com}{hoanglam296@gmail.com}} \\[-0.25cm]
	\normalsize Citigroup
}

\date{\normalsize \today }

\begin{document}
    \maketitle
    \thispagestyle{empty} 
    \vspace*{-0.8cm}
    \begin{abstract}
    \begin{spacing}{1.2}
    \noindent 
We empirically investigate the distributional effects of inflation on workers' unemployment tail risks using instrumental variable quantile regression. We find that supply-driven inflation disproportionately raises unemployment tail risks for cyclically vulnerable workers in both the short and medium term, while demand-driven inflation has differential effects---limited to race and reason for unemployment---only in the medium term. Demand-boosting policies, including monetary policy, can inadvertently widen those disparities through the inflation channel, underscoring the importance of inflation stabilization in promoting equitable growth in the labor market. Our findings could be explained structurally by heterogeneity in experienced inflation and wage inflation expectations.  

    \vspace*{0.5cm}
    
    \noindent \textit{JEL classification:} E24, E32, E66 \\
    \noindent \textit{Keywords:} Quantile regression, Instrumental variable, Unemployment, Inequality, Inflation, Growth at risk, Monetary policy.  
    \end{spacing}
    \end{abstract}

\newpage

\pagebreak
\addtocounter{page}{-1}

\setcounter{page}{1}

\section{Introduction}\label{sec:intro}

The Covid-19 pandemic was an unprecedented shock to the labor market, with substantially uneven effects across different workers and sectors, while driving a dramatic surge in inflation. The rise in wage and price inflation during the pandemic raised concerns about its adverse effects on the labor market and inequality (e.g., \citealp{del2023inflationary}, \citealp{del2024distributionala}). Relatedly, recent studies highlight differential effects of macroeconomic shocks on the behavior and outcomes of workers. \cite{cajner2017racial} claim the labor market outcomes of racial minorities are more cyclically sensitive. \cite{ahn2023} and \cite{Graves2023} demonstrate that job quitters and job losers exhibit opposite cyclical responses in their unemployment incidence in response to cyclical or monetary policy shocks. \cite{PR} highlight that an increase in inflation expectations raises on-the-job search and hence quits. What remains to be explored are the potentially distinct effects of inflation driven by supply and demand shocks on the unemployment risks faced by different groups of workers. This issue is directly related to assessing the real-side costs of high inflation, as well as the costs of running the economy hot, and is therefore central to the design of monetary policy.\footnote{The distributional effects of inflation on labor market outcomes, particularly the risk of joblessness, are central to the connection between the dual mandate in the new monetary policy framework which seeks to promote ``broad-based and inclusive" maximum employment.}   


This paper investigates the effects of inflation by its structural sources on the unemployment risks faced by workers with different socioeconomic attributes, and identifies specific worker groups that are particularly vulnerable to unemployment due to inflation, referred to as the \textit{who-at-risk}. We focus on the \emph{unemployment tail risk}, a concept capturing changes in the unemployment rate that are likely observed during an economic recession, because our focus is on predicting the recession dynamics, particularly the severity of joblessness among different groups during downturns. We follow \cite{kiley2022unemployment} and use quantile regression to measure unemployment tail risk for a given worker group as an increase in its unemployment rate falling within the upper fifth percentile of the group’s conditional distribution of unemployment rate changes.\footnote{The 80th percentile of the change in the unemployment rate is the smallest value such that there is an 80\% or greater probability that the change in the unemployment rate will be less than or equal to that value. For a one-year horizon, the 80th percentile of the change in the unemployment rate is at least 0.75 percentage points, and for a three-year horizon, it is at least 1.9 percentage points (\citealp{kiley2022unemployment}).} 

Quantile regression has emerged as a popular econometric methodology for assessing macroeconomic risks and the predictability of macroeconomic factors in recession forecasts (e.g., \citealp{adrian2019vulnerable}). However, standard reduced-form quantile regression is limited in its ability to identify the effects of structural shocks. Since our focus is the effects of inflation driven by structural sources on unemployment tail risk, we employ \emph{quantile regression with instrumental variables} (henceforth, IVQR)---a recent advancement in econometric analysis that identifies the relationship between an endogenous outcome at a particular quantile and a regressor driven by an exogenous factor (\citealp{chernozhukov2006instrumental}). In this context, we bring a few innovations to the literature on quantile regression and empirical macroeconomics. First, this paper employs instrumental variables to make causal inference within a quantile regression framework and is the first, to our knowledge, to use IVQR to assess macroeconomic risks. On the other hand, most studies on macroeconomic tail risks have used reduced-form quantile regressions. Second, we uncover cross-sectional heterogeneity in unemployment tail risks by estimating the IVQR models for disaggregate data. Cross-sectional heterogeneity has been underexplored in this literature. To the best of our knowledge, this paper is the first to investigate cross-sectional heterogeneity in the effects of inflation on unemployment tail risks and is also the methodologically first one that employs the IVQR to distinguish the effects of inflation by its structural sources.

Assessing the unemployment tail risk, we focus on changes in the unemployment rate rather than its level for several reasons. First, the levels and trends of the unemployment rate vary across different groups, making direct comparisons less reliable for evaluating cyclical dynamics. Considering that the unemployment rates of disadvantaged workers are generally higher and show larger variations than the average, we control for each group’s level of unemployment rate in our model. Second, measuring unemployment tail risk through changes in the unemployment rate aligns conceptually with the approach used to assess risks in real GDP growth within the Growth-at-Risk framework (\citealp{Adrian2019}). This approach also aligns with the widely accepted framework of Okun's law, which links changes in the unemployment rate to real GDP growth. In addition, we consider two forecast horizons for unemployment tail risk: a one-year horizon to evaluate near-term recession risks, and a three-year horizon to capture the medium-term effects of economic shocks and the timeframe for the propagation of monetary policy and macroeconomic risk management.

Distinguishing between the two structural sources of inflation is important, as stabilizing each type requires different policy measures. However, less explored is the extent to which each type of inflation affects the unemployment tail risks faced by different groups of workers. Previous literature has shown that cost-push or supply factors of inflation can cause an economic downturn (\citealp{hamilton2009causes,Gali2011}), which eventually has uneven effects on workers. When facing higher production costs or adverse mark-up shocks---alongside reduced demand due to rising prices---firms are likely to lay off low-productivity workers first when downsizing. A shock of this sort has larger negative effects on workers on the margin of the labor market. The associated increase in uncertainty discourages on-the-job searches and hence reduces the unemployment of job quitters, while raising the unemployment of job losers (\citealp{Clymo}). 

Meanwhile, demand-driven inflation can have opposite effects on unemployment risks. Strong demand reflected in inflation can improve the employment prospects of low-skilled workers (\citealp{AV2023}) and facilitate job-to-job transitions for better worker-job matches (\citealp{Moscarini2023, PR}). As a result, the unemployment of disadvantaged workers declines, but the short-term unemployment of job quitters (reflecting the intervening spell of job searches between the old and new jobs) can rise. Meanwhile, sustained high inflation likely raises wages and causes firms to reduce labor inputs in their production by laying off low-productivity workers. Again, this sort of uncertainty reduces on-the-job searches and the unemployment of voluntary job leavers. Therefore, it's not clear which effect dominates.


We find significantly different effects of inflation depending on its source and substantial cross-sectional heterogeneity in unemployment tail risks across different groups. First, among the various worker attributes considered—including demographic characteristics, education, job status, and reasons for unemployment—cross-sectional heterogeneity is most pronounced by race and by reason for unemployment. 
Second, the supply-driven inflation has more pervasive and immediate distributional effects. Unlike the supply-driven inflation, the demand-driven inflation does not show statistically significant differential effects in the short term. In the medium term, the supply-driven inflation affects all categories considered, disproportionately raising unemployment tail risks for cyclically vulnerable groups---such as racial and ethnic minorities, the less-educated, part-time workers, job losers, and young workers. This finding aligns with the observation that cost-push factors of inflation are recessionary and help predict economic downturns 12 months ahead and over the business-cycle frequency forecasting horizon.

In contrast, the demand-driven inflation widens disparities in the medium term with statistical significance, limited to race and reason for unemployment. Nonetheless, it is important to note that demand-driven inflation significantly raises the unemployment tail risk for Black workers compared to White workers, and lowers the risk for job leavers while raising it for job losers---a pattern and magnitude that are similar to those observed under supply-driven inflation. Despite reflecting strong demand, the persistent feature of demand-driven inflation raises unemployment tail risks, with disproportionately adverse effects on racial and ethnic minorities and job losers, while also discouraging job quits. 

This observation further suggests that expansionary monetary policy can have adverse consequences for disparities in unemployment tail risks. To examine this possibility, we replace demand-driven inflation with an externally identified monetary policy shock from \cite{romer2004new} in our model and find that the portion of inflation driven by expansionary monetary policy exhibits differential effects in unemployment tail risks similar to those of demand-driven inflation. This portion of inflation raises the tail risk for Black workers while not affecting the risk of Whites in both the short and medium term. Its differential effects in the medium term are somewhat muted by reason for unemployment, but qualitatively similar to those of demand-driven inflation. All told, a demand-boosting policy---if inflationary---exacerbates disparities in labor market outcomes rather than improving them.

Furthermore, we find that wage setting and wage inflation expectations may be important underlying factors behind the adverse effects of both supply- and demand-driven inflation on disparities in unemployment tail risks. In the short term, wage inflation resembles supply-driven inflation, while in the medium term, it mirrors the cross-sectional effects of demand-driven inflation---particularly by race and reason for unemployment. These results highlight the interplay between wage setting and price inflation in shaping the distributional effects on unemployment tail risks through both supply and demand channels.

What structural mechanism explains this observed link? First, wage inflation itself can act as a cost-push factor (e.g., through changes in minimum wages) and simultaneously reflect the demand factor of inflation, as suggested by the relatively steep wage Phillips curve (\citealp{Gali_Gambetti2019}). Second, both supply and demand sources of inflation can raise wage inflation expectations, particularly more so for racial and ethnic minorities, in response to supply-driven inflation
(\citealp{Lee2022,Orchard2022}).\footnote{\cite{Lee2022} finds that racial and ethnic minorities face higher inflation rates than Whites. \cite{Orchard2022} shows that low-income workers are exposed more to prices of necessities that are largely determined by energy and food prices.} The effects of supply-driven inflation tend to emerge more quickly in the short term, as firms facing rising input costs are more likely to cut jobs held by low-skilled, low-productivity workers—who are disproportionately from racial and ethnic minority groups—rather than retain them. In addition, these workers---exposed more to necessity prices largely determined by food and energy costs (\citealp{Orchard2022})---are likely to demand higher wage increases, which further raises their unemployment risks. In contrast, the impact of demand-driven inflation unfolds more gradually. While firms may initially retain existing workers to meet increased demand despite the increased wage inflation due to strong labor demand, sustained upward pressure on labor costs can incentivize the adoption of labor-saving technologies, ultimately displacing workers with low productivity or skills. In this way, both types of inflation---in connection with wage inflation---generate differential effects on unemployment tail risks. Importantly, heterogeneity in experienced inflation and wage inflation expectations can be an important channel through which inflation generates distributional effects on unemployment tail risks. 


Lastly, we examine the extent to which oil supply shocks account for the distributional effects of supply-driven inflation and the unemployment tail risks. For this, we employ the oil supply news shocks from \cite{kanzig2021macroeconomic} as the instruments. Supply-driven inflation and the portion of inflation attributed to oil supply news shocks exhibit similar distributional effects, with some notable differences. Specifically, inflation driven by oil supply shocks generates more muted cross-sectional heterogeneity across racial groups, while producing larger differential effects by reason for unemployment---particularly by reducing the unemployment tail risk for job leavers more than general supply-driven inflation. This finding, along with the distributional effects of monetary policy, suggests that the supply and demand drivers of inflation are not confined to specific structural shocks.\footnote{Other events, such as terrorist attacks, financial crises, pandemics, and supply chain disruptions, can also significantly influence inflation dynamics.} In addition, notably rising inflation driven by oil supply shocks discourages on-the-job search more than general supply-driven inflation.

Our empirical findings have significant implications for both monetary policy and demand-management policies. First, the statistically significant distributional effects of supply- and demand-driven inflation suggest that inflation stabilization can also help mitigate disparities in unemployment tail risks---a less-recognized consequence of inflation stabilization. Second, the medium-term distributional effects of demand-driven inflation, along with the distributional pass-through of monetary policy, suggest that demand-boosting policies aimed at reducing employment shortfalls among disadvantaged workers may inadvertently increase their unemployment risks through the inflation channel, thereby exacerbating labor market inequality. All told, maintaining inflation stability is essential when implementing policies aimed at promoting equitable growth in the labor market. 


The plan of the paper is as follows. Section \ref{sec:literature_review} provides a detailed literature review focusing on the contribution of this paper. Section \ref{sec_motivation} motivates the need to distinguish between supply- and demand-driven inflation when assessing unemployment risks, as well as the cross-sectional heterogeneity in those risks. Section \ref{sec:model} introduces the model and describes the data. Section \ref{sec:results} presents the empirical findings on aggregated unemployment risks. Section \ref{sec:robustness} delves deeper into the structural sources of inflation. Section \ref{sec:conclusion} concludes.

\section{Literature Review} \label{sec:literature_review}

This paper makes several unique contributions to the literature on inequality in macroeconomics and that on quantile regression by bridging the gap between the two. Based on the IVQR, we highlight the importance of the labor market through which two types of inflationary shocks exert discernible distributional effects. This novel empirical insight can offer a guiding principle for the development of structural macroeconomic models.

To begin with, this paper contributes to the literature on quantile regressions by incorporating cross-sectional heterogeneity and applying this method to the study of inequality, providing a novel empirical application. 
Quantile regression has become popular for the assessment of macroeconomic risks. The method has been applied to evaluate the tail risk of aggregate outcomes such as GDP growth (\citealp{adrian2019vulnerable}), inflation (\citealp{lopez2024inflation}), unemployment (\citealp{kiley2022unemployment}), bank performance (\citealp{nguyen2023variable}), among others. Our research expands the literature on quantile regressions in two ways. First, most of the studies on macroeconomic tail risks have used reduced-form quantile regressions (\citealp{koenker1978regression}), whereas our paper employs instrumental variables to make causal inferences in the IVQR (e.g., \citealp{chernozhukov2008instrumental}; \citealp{chernozhukov2017instrumental}). Second, aggregate risk models conceal the heterogeneous risk exposures of economic agents. Despite the growing importance of heterogeneity in macroeconomic modeling, cross-sectional heterogeneity in the risks has been underexplored, which our research fills the gap.

In addition, this paper makes a unique contribution to the literature on inequality in the context of the distributional effects of inflation on labor-market disparities. Most studies focus separately on either the effects of inflation on inequality or labor-market disparities. Our paper is the first to link the two, highlighting the central role of the labor market. For example, \cite{Doepke2006} find that the primary losers from inflation are wealthy, older households who hold substantial nominal assets, whereas the main beneficiaries are younger, middle-class households with fixed-rate mortgages. \cite{Orchard2022} studies cyclical variation in inflation rates by income level, finding that low-income households experience higher consumption price inflation during economic recessions than do high-income households. \cite{FLR2022} show that stock returns are negatively correlated with core inflation and conclude that holding stocks offers little scope to hedge against inflation risk.

In addition, emerging studies focus on the heterogeneous effects of monetary policy through the inflation channel. \cite{lee2021minority} presents evidence that accommodating monetary policy helps reduce the racial disparity between Black and White workers as long as inflation expectations remain anchored, which aligns with the main implication of this paper. \cite{del2023inflationary} find regressive effects of cost-push inflation and progressive effects of demand-driven inflation, and claim the importance of considering various aspects of the budget constraint to correctly assess the welfare effects of inflation. Specifically, \citeauthor{del2023inflationary} highlight the asset price channels in explaining the welfare differences between demand-driven and supply-driven inflation. Following an oil-price shock, asset prices decline, and middle-aged households with college education can acquire more equity. On the other hand, an increase in asset prices following a monetary expansion will have an opposite effect. Thus, the authors argue that oil-supply shock is regressive, while expansionary monetary policy is progressive.

Our research is also related to the literature on labor market disparities. Researchers have employed conventional measures of labor market outcomes such as unemployment, job losses and findings, wages, and so on (e.g., \citealp{jefferson2008educational}; \citealp{hoynes2012suffers}; \citealp{cajner2017racial}; \citealp{doniger2021ways}). For example, \cite{doniger2021ways} focus on wages, and \cite{cajner2017racial} consider labor market flows by demographic characteristics. \cite{hoynes2012suffers} find that the people who suffer the most during recessions are men, Blacks, Hispanics, youth, and those with lower education levels. In contrast to these studies, we show that unemployment tail risks can be a new metric for the comprehensive assessment of disparities in labor market outcomes.

This paper is methodologically similar to \cite{kiley2022unemployment} and shares a similar research focus with \cite{del2023inflationary}. \cite{kiley2022unemployment} employs quantile regressions to investigate the role of financial conditions in shaping aggregate unemployment risks. However, our study diverges from \cite{kiley2022unemployment} in two key ways. First, we use instrumental variables quantile regression (IVQR) to focus on the causal effects of inflation on unemployment risks based on its structural sources. Second, we expand the scope to examine unemployment risks across disaggregated worker groups, enabling us to investigate the distributional consequences of inflation.

Similarly, our paper also aligns with \cite{del2023inflationary} in differentiating inflation by its structural sources and exploring their distributional effects. However, there are important differences between the two studies. \citeauthor{del2023inflationary} distinguish between inflation types using oil supply shocks and monetary policy shocks within a structural VAR model, whereas we utilize supply- and demand-driven inflation measures from \cite{shapiro2022much} within the IVQR framework. We further demonstrate that supply and demand factors are not limited to oil supply shocks and monetary policy shocks, respectively. Importantly, while \citeauthor{del2023inflationary} focus on household welfare inequality through the budget constraint, our primary emphasis is on disparities within the labor market. Notably, we do not observe regressive or progressive effects of oil supply shocks and monetary policy shocks, respectively, on unemployment tail risks. This finding is consistent with \cite{del2023inflationary}, which argues that the effects of structural shocks on inequality do not operate through the labor market channel. Our study underscores unemployment tail risks as a novel channel through which inflation influences inequality, differentiating it from \cite{del2023inflationary}.

\section{Inflation and Heterogeneity in Unemployment Risk: A Primer} \label{sec_motivation}

\cite{kiley2022unemployment} employs a quantile regression and assesses the unemployment tail risk as defined as an increase in unemployment rate falling in the upper fifth percentile of the distribution of unemployment rate changes. We estimate the three-year-ahead unemployment tail risk (three-year changes in the unemployment rate) of Whites and Blacks based on \citeauthor{kiley2022unemployment}'s model.\footnote{We extend the model from \cite{kiley2022unemployment} for our empirical analyses. Section A of the online appendix outlines the methodology for the reduced-form quantile regression.}

\begin{figure}[t]
\caption{Unemployment-at-Risk: Blacks v. Whites}\label{figure:TS_PCE_UAR}
  \begin{center}
	\includegraphics[width=\textwidth]{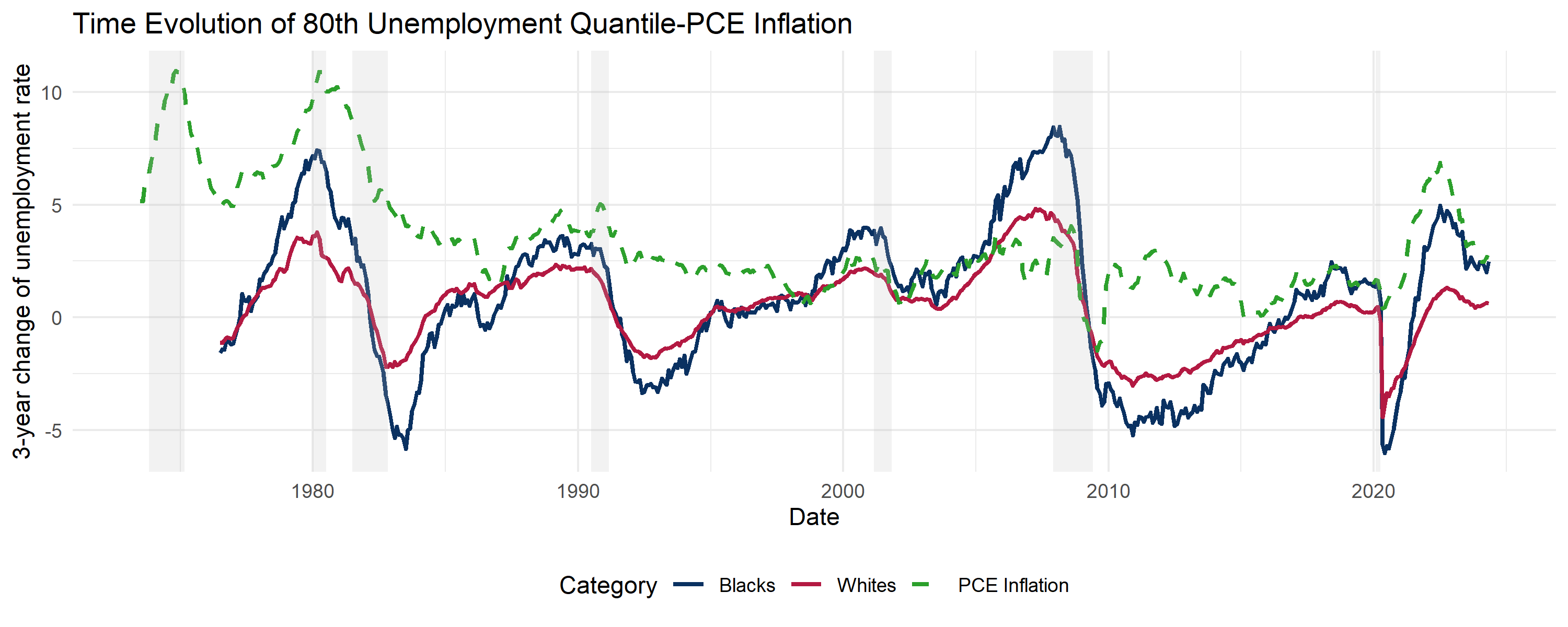}    
  \end{center}  
{\footnotesize \textbf{Notes to figure}. This figure shows the 80\% quantiles of the three-year-ahead changes in unemployment rates for Blacks and Whites, alongside the year-over-year change in the PCE chain-type price index. \textcolor{black}{The distribution is estimated using reduced-form quantile regressions of \cite{kiley2022unemployment} where the PCE price inflation is one of the predictors.} \textbf{Source:} Authors' calculation.} 
\end{figure}

Figure \ref{figure:TS_PCE_UAR} displays the unemployment tail risk of Whites (red line) and Blacks (navy line) along with the PCE price inflation (dashed green line). There are a few important observations. First, the tail risks for Blacks and Whites differ significantly, with Blacks’ unemployment risks showing more cyclical behavior. Second, the PCE price inflation is closely linked to these tail risks. Despite overall unemployment remaining low, Blacks' unemployment risk spiked during the high inflation period following the COVID-19 recession, reaching unprecedented levels, exceeded only by those during the 1980 and 2008 recessions. In contrast, Whites' unemployment risk showed only a slight increase from its pre-/COVID-19 levels. This observation suggests that price inflation has larger power to predict Blacks' unemployment tail risk. 

\begin{figure}[t]
\caption{Historical Unemployment Rate Distribution: Blacks v. Whites}\label{figure:UR_Historical}
	\begin{center}
	\begin{subfigure}[b]{0.49\textwidth}
         \centering
         Panel A. March 1980 \\
         \includegraphics[width=1\textwidth, height = 0.6\textwidth]{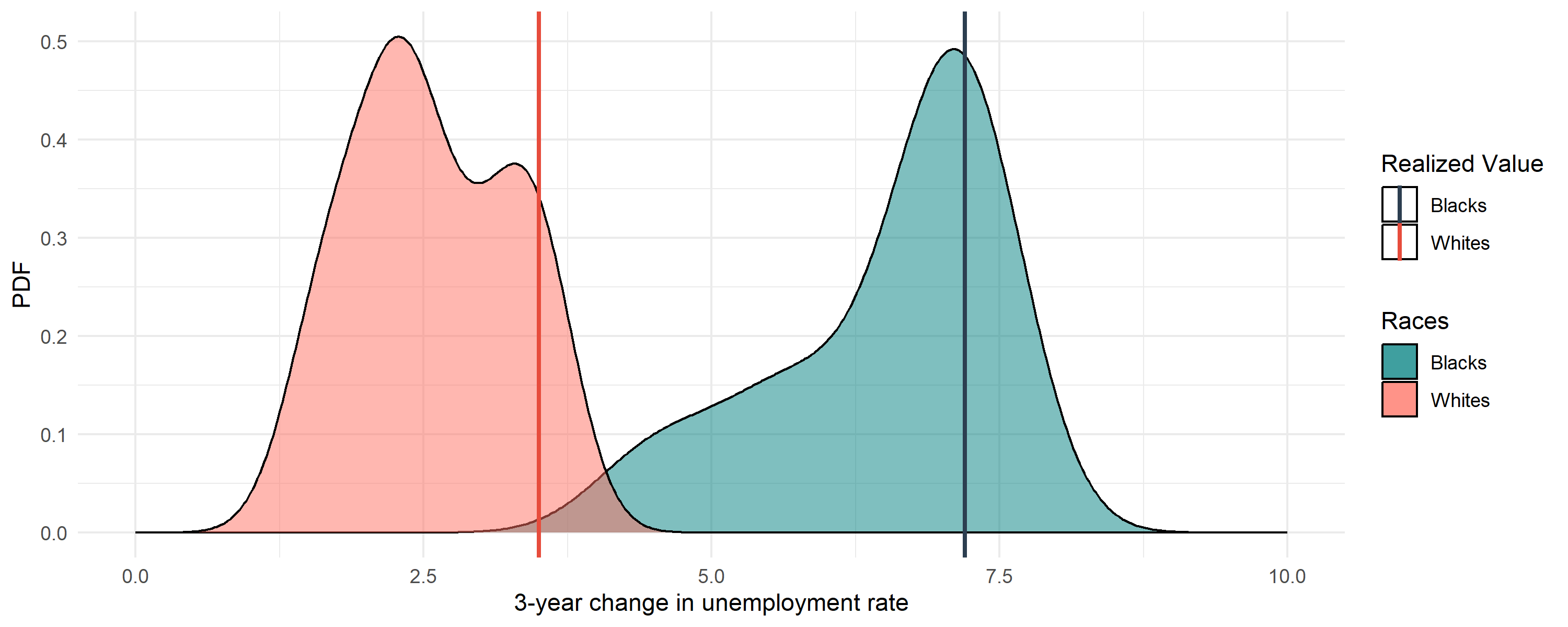}
     \end{subfigure}
     \hfill
     \begin{subfigure}[b]{0.49\textwidth}
         \centering
         Panel B. May 1989
         \includegraphics[width=1\textwidth, height = 0.6\textwidth]{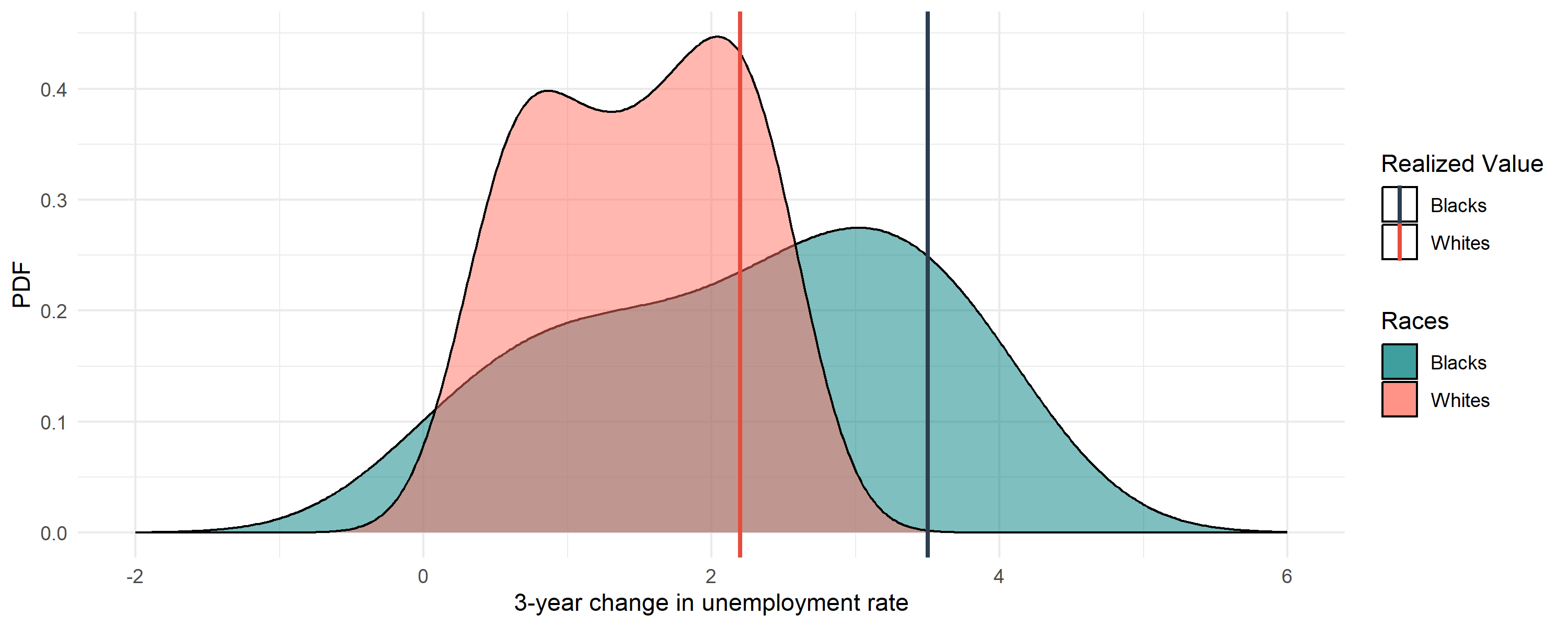}
     \end{subfigure}      
     \end{center} 
     \begin{center}
	\begin{subfigure}[b]{0.49\textwidth}
         \centering 
         Panel C. July 2008
         \includegraphics[width=1\textwidth, height = 0.6\textwidth]{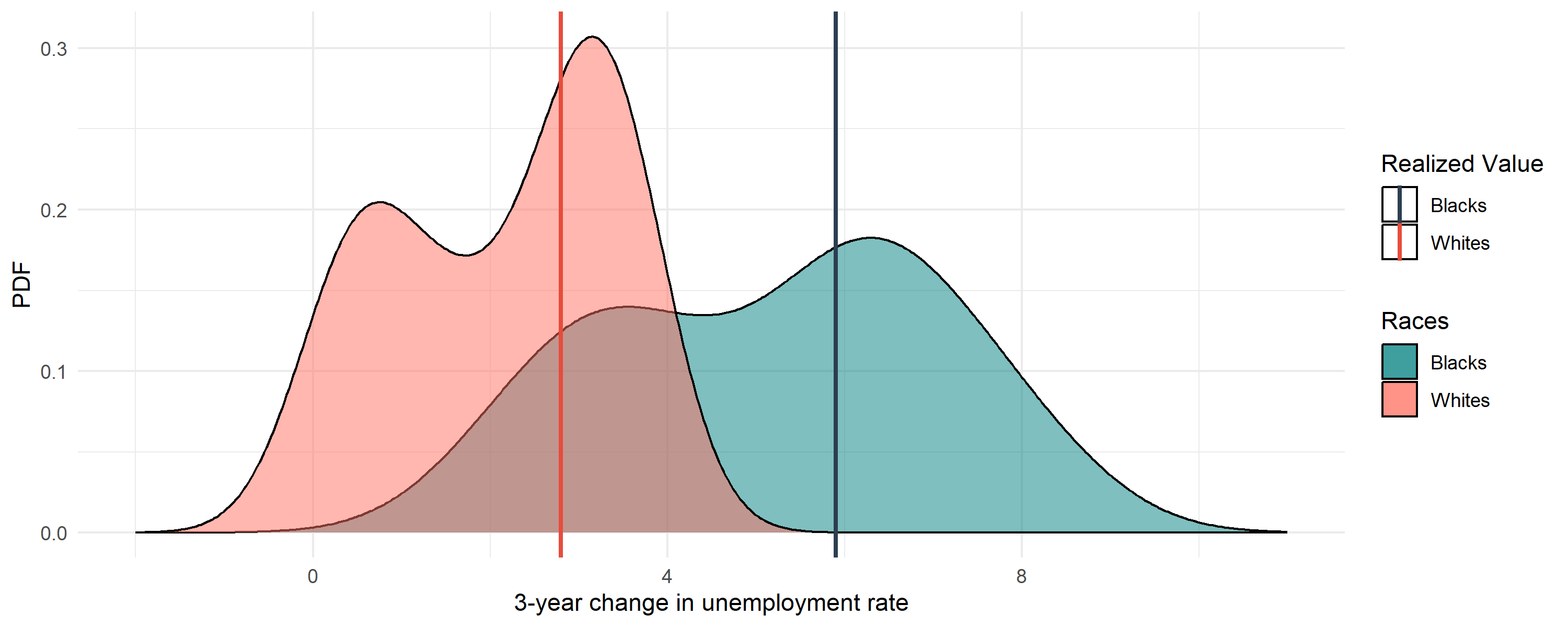}
     \end{subfigure}
     \hfill
     \begin{subfigure}[b]{0.49\textwidth}
         \centering
         Panel D. June 2022
         \includegraphics[width=1\textwidth, height = 0.6\textwidth]{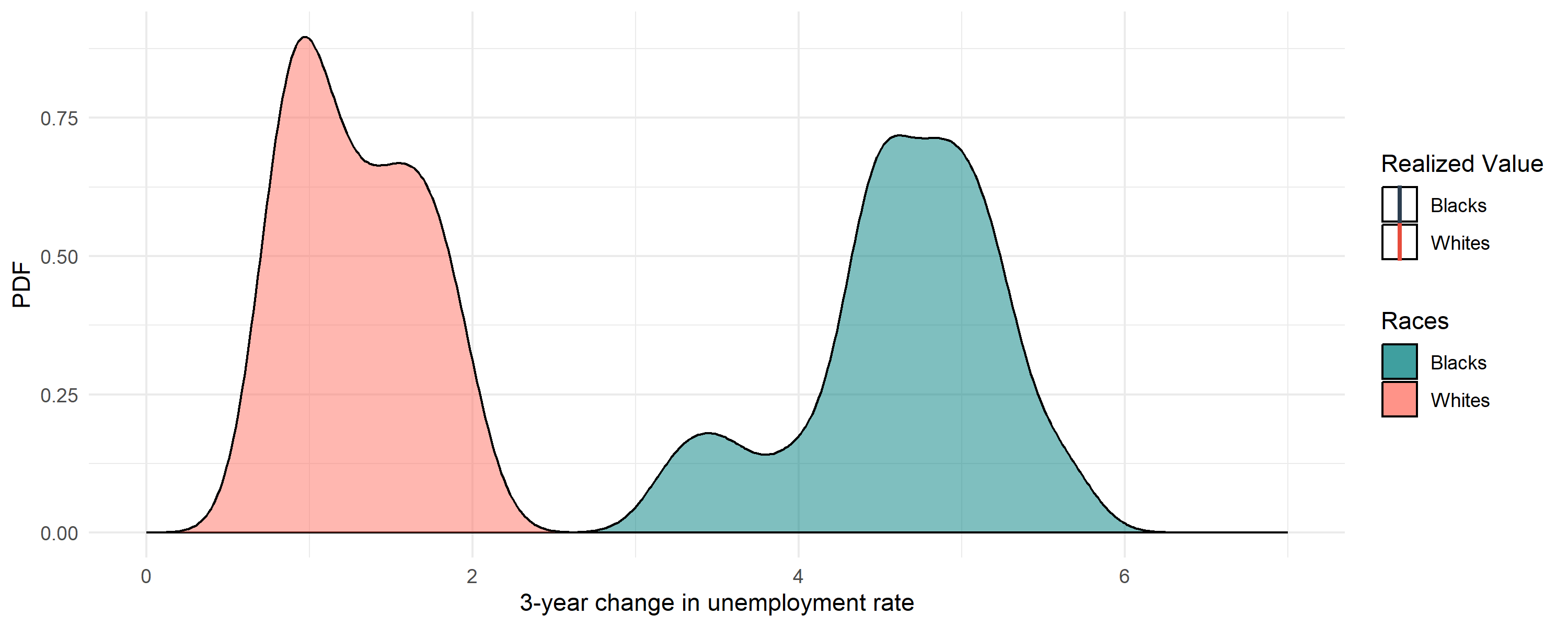}
     \end{subfigure}      
     \end{center} 
{\footnotesize \textbf{Notes to figure}. This figure shows the predictive density of the 3-year-ahead unemployment rates for Blacks and Whites, along with their realized values. \textcolor{black}{The distribution is estimated using reduced-form quantile regressions of \cite{kiley2022unemployment} where the PCE price inflation is one of the predictors.}  \textbf{Source:} Authors' calculation.} 
\end{figure}

Figure \ref{figure:UR_Historical} presents the 3-year-ahead predictive density of changes in Black and White unemployment rates, estimated from the 10th, 15th, 20th, and other quantiles up to the 90th, to cover the entire distribution. We focus on episodes of high inflation over the past 40 years: March 1980, when inflation peaked at 11\% (Panel A); May 1989, following a steady rise to 5\% (Panel B); July 2008, when inflation reached 4\% during the Great Recession (Panel C); and June 2022, when inflation peaked at 7\% (Panel D). The vertical lines represent the realized values for each group.\footnote{Panel D does not have vertical lines, since we do not have the data for Jun 2025 yet.} Panels A--C demonstrate that the mode of each group's distribution closely aligns with its corresponding vertical line, indicating that the density forecasts are reasonably accurate and effectively capture the associated risks.

Across all episodes, significant heterogeneity in unemployment risk distributions is observed. Notably, the distributions for Blacks and Whites are similar in the 1980 and 2022 recessions, showing 
significant dispersion with minimal overlap between the two racial groups. Note that both periods were marked by high price inflation driven by supply-side factors---such as oil price shocks in the early 1980s and supply bottlenecks and labor shortages during the pandemic. These periods occurred at the peak of economic expansion and on the brink of recessions, indicating that overall demand was also strong (and hence followed by monetary tightening). In contrast, oil prices were relatively low and stable in 1989, while in 2008 the economy faced a severe demand-driven recession despite elevated oil prices. As a result, inflation was only moderately high in both periods, and the distributions for White and Black workers showed greater overlap, with a smaller gap in unemployment tail risks between the two groups. 

This observation carries two important implications. First, high inflation can exacerbate disparities in unemployment risk across different groups, with racial disparities being a prominent example. Second, the underlying supply and demand forces driving inflation may have distinct effects on the tail risks of unemployment among various groups of workers. Relatedly, oil shocks can be recessionary, and cyclical sensitivity varies across different segments of the labor force (\citealp{cajner2017racial}). While some studies, such as \cite{AV2023}, suggest that prolonged expansionary monetary policy can draw marginal workers into employment---reducing joblessness without significantly increasing inflation—there is little evidence on how demand-driven inflation affects unemployment tail risks. All told, distinguishing between supply- and demand-driven inflation is crucial for a comprehensive analysis of unemployment tail risks across different groups of workers. 

\section{Model}\label{sec:model}

This section discusses the analytical framework. Section \ref{sec_URisk_IVQR} introduces the IVQR methodology and section \ref{ssec:estimation} discusses the estimation method. Section \ref{sec_data} describes the data. And Section \ref{sec_2sls_ivqr} explains why IVQR is needed instead of the linear IV model for assessing unemployment tail risks. 

\subsection{Structural Factors of Unemployment Risks and the IVQR} \label{sec_URisk_IVQR}

Our focus is effects on unemployment risks of inflation by its structural sources. An approach similar to two-stage least squares with an instrumental variable is employed within a quantile regression. This methodology is known as the IVQR model. 

To illustrate, suppose that we investigate the effects of inflation driven by structural shocks on the unemployment risk. The dependent variable, $\triangle y_{t+h}$, is the $h$-period-ahead change in the unemployment rate of group $i$, and we suppress $i$ for notational convenience. Let $\mathbf{z}_t$ denote the instrument of structural shock and $\mathbf{d}_{t}$ denote the endogenous covariates, which are just inflation in our case. Also, let $\mathbf{x}_t$ denote the set of macro controls. If we simply care about the effects of inflation on the conditional mean of unemployment, we can use the linear IV estimator on the following regression:
\begin{equation}
   \triangle y_{t+h}=\mathbf{d}_{t}^{\prime}\bm{\alpha}+\mathbf{x}_{t}^{\prime}\bm{\beta}+e_{t+h},
\end{equation}
which can be estimated from the following conditional moment restriction:
\begin{equation}
    E\left(\triangle y_{t+h} - \mathbf{d}_{t}^{\prime}\bm{\alpha}-\mathbf{x}_{t}^{\prime}\bm{\beta} |\mathbf{x}_{t},\mathbf{z}_{t}\right)=0.
\end{equation}

Now, since we are interested in the heterogeneous effects across quantiles, we need to estimate a quantile regression model as follows:
\begin{equation}
   \triangle y_{t+h}=\mathbf{d}_{t}^{\prime}\bm{\alpha}(\chi)+\mathbf{x}_{t}^{\prime}\bm{\beta}(\chi),
\end{equation}
where ${\bm \alpha}(\chi)$ and ${\bm \beta}(\chi)$  are the coefficients of $\mathbf{d}_{t}$ and $\mathbf{x}_{t}$ for the $\chi$ quantile. 

Under some regularity conditions, the IVQR model for quantile $\chi$ satisfies the following conditional probability 
\begin{equation}
    \Pr\left(\triangle y_{t+h} \le \mathbf{d}_{t}^{\prime}\bm{\alpha}(\chi)+\mathbf{x}_{t}^{\prime}\bm{\beta}(\chi)|\mathbf{x}_{t},\mathbf{z}_{t} \right)=\chi,
    \label{eq:IVQR_conditional}
\end{equation}

\noindent which is equivalent to the following conditional moment restriction
\begin{equation}
    E\left(\left[\chi-I\left\{\triangle y_{t+h} \le \mathbf{d}_{t}^{\prime}\bm{\alpha}(\chi)+\mathbf{x}_{t}^{\prime}\bm{\beta}(\chi) \right\}\right]|\mathbf{x}_{t},\mathbf{z}_{t} \right)=0. 
    \label{eq:IVQR_conditional_moment}
\end{equation}

\noindent Similar to the linear IV case, equation \eqref{eq:IVQR_conditional_moment} suggests that any measurable function of $\mathbf{x}_{t}$ and $\mathbf{z}_{t}$ can be a valid instrument for the IVQR model. Thus, it implies the following unconditional moment condition:
\begin{equation}
    E\left(\left[\chi-I\left\{\triangle y_{t+h} \le \mathbf{d}_{t}^{\prime}\bm{\alpha}(\chi)+\mathbf{x}_{t}^{\prime}\bm{\beta}(\chi) \right\}\right]\Psi_{t}\right)=0, 
    \label{eq:IVQR_moment}
\end{equation}
where $\Psi_{t}$, a function of $\mathbf{x}_t$ and $\mathbf{z}_t$,  serves as an instrument for $\mathbf{d}_t$. Following the random-coefficient representation of quantile regression, we assume that $\chi|\mathbf{x}_{t},\mathbf{z}_{t} \sim \text{Uniform}(0,1)$. 

Equation \eqref{eq:IVQR_moment} is the quantile analog of the moment condition in the linear IV model. The only difference is that the objective function based on \eqref{eq:IVQR_moment} is nonconvex and nonsmooth due to the indicator function. Although the intuition is straightforward, the estimation of the IVQR model is numerically involved. 

One may conjecture obtaining the fitted predictors from an external first-stage regression and then plugging the fitted predictors into the regular quantile regression which is likely to be computationally less involved. In fact, \cite{adrian2022term} adopt such an approach. However, the quantile analog of the two-stage linear regression (i.e. the two-stage quantile regression estimator) is not a proper choice for our case where the structural factors' effects are heterogeneous across quantiles. In such a case, \cite{chernozhukov2006instrumental} note that the two-stage quantile regression estimator produces inconsistent estimates.\footnote{See footnote 1 of \cite{chernozhukov2006instrumental}.} For this reason, we use the IVQR. 

\subsection{Estimation} \label{ssec:estimation}

Estimating an IVQR model is numerically involving. For the estimation, we first use the inverse quantile regression estimator of \cite{chernozhukov2006instrumental} as a benchmark, and we switch to the smooth estimator of \cite{kaplan2017smoothed} when the first estimator runs into numerical problem.\footnote{We use the STATA commands \textit{ivqregress iqr} and \textit{ivqregress smooth} to implement the inverse quantile regression estimator and the smooth estimator, respectively. For more details on the estimation, interested readers can consult \cite{stata2023}.}     

\cite{chernozhukov2006instrumental} propose to use a grid search. To understand their method, note that equation \eqref{eq:IVQR_conditional} can be rewritten into the following:
\begin{equation}
    \Pr\left(\triangle y_{t+h} - \mathbf{d}^{\prime}_{t}\bm{\alpha}(\chi) \le \mathbf{x}_{t}^{\prime}\bm{\beta}(\chi)+\mathbf{z}_{t}^{\prime}*0|\mathbf{x}_{t},\mathbf{z}_{t} \right)=\chi. 
\label{eq:IVQR}
\end{equation} 
\noindent 

\noindent Intuitively, equation \eqref{eq:IVQR} suggests that if $\mathbf{d}_{t}$ adequately captures effects of $\bm{z}_t$ and the true values of $\bm{\alpha}$ are known, $\bm{z}_t$ should not have explanatory power in predicting the unemployment risk at $\chi$ percentile. Thus, $\triangle y_{t+h}-\mathbf{d}_{t}^{\prime}\bm{\alpha}(\chi)$ conditional on $\mathbf{x}_{t}$ and $\mathbf{z}_{t}$ should produce coefficients close to zeros on $\mathbf{z}_{t}$. Leveraging this feature, the model is estimated with a grid search on possible values of $\bm{\alpha}$. The estimation involves the following steps. First, calculate $\hat{\mathbf{d}}_{t}$ from the first-stage regressions where $\mathbf{d}_{t}$ is regressed on $\mathbf{x}_{t}$ and $\mathbf{z}_{t}$. Second, define a grid ${\bm{\alpha}_{1},\dots,\bm{\alpha}_{K}}$. Third, for each $\bm{\alpha}_{k}$ in that grid, estimate the quantile regression where $\triangle y_{t+h}-\mathbf{d}^{\prime}_{t}\bm{\alpha}_{k}(\tau)$ is the dependent variable and the regressors are $\mathbf{x}_{t}$ and $\hat{\mathbf{d}}_{t}$. Fourth, find the solution, $\bm{\alpha}^{*}$ that makes the coefficients of $\mathbf{\hat{d}}_{t}$ closest to \textbf{0} by using the Wald statistics.

On the other hand, \cite{kaplan2017smoothed} suggest to approximate the moment condition \eqref{eq:IVQR_moment} with 
\begin{equation}
    E\left(\left[\tilde{I}\left\{\dfrac{1}{h}[\triangle y_{t+h} - \mathbf{d}_{t}^{\prime}\bm{\alpha}(\chi)-\mathbf{x}_{t}^{\prime}\bm{\beta}(\chi)]-\chi \right\}\right]\Psi_t\right)=0, 
    \label{eq:IVQR_KaplanSun}
\end{equation}

\noindent where $\tilde{I}(.)$ is a piecewise linear function and $h$ is the theoretical optimal bandwidth.\footnote{In particular, $\tilde{I}(v)=1$ for $v \le -1$, $\tilde{I}(v)=0$ for $v \geq 1$, and $\tilde{I}(v)=(1-v)/2$ for $-1<v<1$. For more details, interested readers can consult \cite{kaplan2022smoothed}.} Then, the system of smooth, nonlinear equations can be solved with the General Method of Moments (GMM). One advantage of this estimator is that it is fast and can handle multiple instruments. Finally, for confidence interval (CI), we report the robust standard errors for both estimators.  

\subsection{Variables and Data} \label{sec_data}

This section discusses the data used in this paper. Our main interest is the unemployment risks of disaggregate groups. Thus, besides the aggregate unemployment rate, we consider unemployment rate data of six different categories: (1) gender, (2) race, (3) age, (4) education, (5) job status (full-time/part-time), and (6) reason for unemployment. 

As in \cite{adrian2019vulnerable} and \cite{kiley2022unemployment}, financial variables have proved their predictive power of near-term recession and have often been employed as predictors of quantile regressions. We further examine and show that these variables have power in predicting the medium term. The financial regressors are (1) the non-financial leverage National Financial Condition Index that captures movement in the credit cycle; (2) the adjusted NFCI index (ANFCI) that captures general financial condition; (3) the term spread calculated as the difference between 10-Year and 2-Year constant maturity Treasury.\footnote{\cite{Adrian2019} claim that the National Financial Conditions Index is a measure that is linked to the tail risks of real activity. \cite{kiley2022unemployment} mentions that the corporate bond spread shows similar predictive power, and has a longer sample period than the National Financial Conditions Index.} In addition to these financial variables, we also include (4) the current unemployment rate of the corresponding group and (5) the year-over-year change in the headline PCE price index.\footnote{\cite{kiley2022unemployment} finds that the current unemployment rate has a negative correlation with the future unemployment rate, because of its mean-reverting feature, and that the price inflation also raises unemployment risks.} These macro controls are captured in the vector $\mathbf{x}_{t}$. 


We use the estimates of supply-driven and demand-driven inflation identified by sign restrictions from \cite{shapiro2022much} as the instruments. Shapiro identifies the supply and demand shocks based on the signs of movements in quantity and price and employs a model to recover the portion of inflation attributable to these structural shocks.\footnote{\cite{shapiro2022much} identifies a demand shock and a supply shock based on the signs of movements in price and quantity: A demand shock will move price and quantity in the same direction while a supply shock will move them in the opposite direction. \citeauthor{shapiro2022much} uses the sign of the residuals from 10-year rolling regressions on prices and quantities of more than 100 goods and services categories of the PCE price index in order to classify whether a category has experienced \textit{at least} a demand shock or \textit{at least} a supply shock in a given month. Then, the supply-driven and demand-driven inflation are calculated as the weighted average of those categories.} Since \citeauthor{shapiro2022much}'s estimates are likely to be observed with measurement errors, we use the estimates as instruments for the true supply-driven and demand-driven inflation. Specifically, we employ \citeauthor{shapiro2022much}'s supply-driven inflation as $\mathbf{z}_t$ to analyze effects of supply-driven inflation on the unemployment risks, and separately employ \citeauthor{shapiro2022much}'s demand-driven inflation as $\mathbf{z}_t$ to analyze effects of demand-driven inflation on the unemployment risks.


For robustness checks, we further consider various other instruments, including the oil supply news shocks of \cite{kanzig2021macroeconomic}, and narrative-based monetary policy shocks of \cite{romer2004new} extended by \cite{wieland2020financial}.\footnote{We further consider high-frequency monetary policy shocks of \cite{bu2021unified} and labor supply and demand shocks estimated from the sign-restricted SVAR of \cite{baumeister2015sign}. See the appendix for more details.}

\begin{table}[h!]
	\caption{Summary Statistics} 
 \begin{center}
	\scalebox{0.85}{
	\begin{tabular}{l|l|c|c|c|c|c|c|c} \hline \hline 
\textbf{Categories} & \textbf{Variable Names} & \textbf{First} & \textbf{Last} & \textbf{Obs} & \textbf{Mean} & \textbf{SD} & \textbf{Min} & \textbf{Max} \\ \hline\hline
Aggregate & Unemployment Rate & 1948M1 & 2024M4 & 916 & 5.7 & 1.7 & 2.5 & 14.8 \\ \hline
\multirow{4}{*}{Gender} & Women & 1948M1 & 2024M4 & 916 & 6.0 & 1.6 & 2.7 & 16.2 \\
 & Men & 1948M1 & 2024M4 & 916 & 5.6 & 1.9 & 2.3 & 13.5 \\
 & Married Women & 1955M1 & 2024M4 & 832 & 4.5 & 1.4 & 1.9 & 13.1 \\
 & Married Men & 1955M1 & 2024M4 & 832 & 3.4 & 1.3 & 1.4 & 9.6 \\ \hline\
\multirow{4}{*}{Race} & Black or African American & 1972M1 & 2024M4 & 628 & 11.5 & 3.3 & 4.8 & 21.2 \\
 & Hispanic or Latino & 1973M3 & 2024M4 & 614 & 8.5 & 2.6 & 3.9 & 18.9 \\
 & White & 1954M1 & 2024M4 & 844 & 5.2 & 1.5 & 3.0 & 14.2 \\
 & Asian & 2000M1 & 2024M4 & 292 & 4.7 & 1.9 & 2.0 & 14.8 \\ \hline
\multirow{8}{*}{Age} & Men: 16-19 Yrs & 1948M1 & 2024M4 & 916 & 16.8 & 4.5 & 6.4 & 30.7 \\
 & Women: 16-19 Yrs & 1948M1 & 2024M4 & 916 & 15.1 & 3.7 & 5.8 & 37.5 \\
 & Men: 20-24 Yrs & 1948M1 & 2024M4 & 916 & 9.7 & 3.1 & 3.2 & 23.1 \\
 & Women: 20-24 Yrs & 1948M1 & 2024M4 & 916 & 8.7 & 2.4 & 3.0 & 27.9 \\
 & Women: 25-54 Yrs & 1948M1 & 2024M4 & 916 & 5.0 & 1.4 & 2.2 & 13.7 \\
 & Men: 25-54 Yrs & 1948M1 & 2024M4 & 916 & 4.4 & 1.7 & 1.5 & 12.1 \\
 & Men: 55 Yrs \& Over & 1948M1 & 2024M4 & 916 & 3.9 & 1.3 & 1.5 & 12.1 \\
 & Women: 55 Yrs \& Over & 1948M1 & 2024M4 & 916 & 3.6 & 1.2 & 1.6 & 15.3 \\ \hline
\multirow{4}{*}{Education} & Less than High School Diploma & 1992M1 & 2024M4 & 388 & 8.8 & 2.8 & 4.3 & 21.3 \\
 & High School Graduate, No   College & 1992M1 & 2024M4 & 388 & 5.6 & 2.1 & 3.2 & 17.7 \\
 & Some College, Less than   Bachelor Deg & 1992M1 & 2024M4 & 388 & 4.6 & 1.8 & 2.4 & 15.6 \\
 & Bachelor Degree \& Higher & 1992M1 & 2024M4 & 388 & 2.8 & 1.0 & 1.5 & 8.4 \\ \hline
\multirow{4}{*}{Job Status} & Part-Time Workers: Men & 1968M1 & 2024M4 & 676 & 7.7 & 1.9 & 3.7 & 22.7 \\
 & Full-Time Workers: Women & 1968M1 & 2024M4 & 676 & 6.3 & 1.9 & 3.2 & 13.6 \\
 & Full-Time Workers: Men & 1968M1 & 2024M4 & 676 & 5.8 & 2.0 & 1.8 & 12.3 \\
 & Part-Time Workers: Women & 1968M1 & 2024M4 & 676 & 5.6 & 1.6 & 3.3 & 25.4 \\ \hline
\multirow{4}{*}{Reason of U} & Job Leavers {[}Quit Job{]} & 1967M1 & 2024M4 & 688 & 11.5 & 2.6 & 2.4 & 18.1 \\
 & Job Losers / Finished   Temporary Job & 1967M1 & 2024M4 & 688 & 3.1 & 1.3 & 1.2 & 13.2 \\
  & Labor Force Reentrants & 1967M1 & 2024M4 & 688 & 1.7 & 0.3 & 0.9 & 2.4 \\
 & Labor Force New Entrants & 1967M1 & 2024M4 & 688 & 0.6 & 0.2 & 0.2 & 1.3 \\\hline
\multirow{4}{*}{Instruments} & Oil News Shock & 1975M1 & 2023M6 & 582 & 0.0 & 0.6 & -2.9 & 2.0 \\
 & \citeauthor{romer2004new} Monetary Policy Shock & 1969M1 & 2007M12 & 468 & 0.0 & 0.3 & -3.2 & 1.9 \\
 & \cite{bu2021unified} Monetary Policy Shock & 1994M1 & 2023M9 & 357 & 0.0 & 0.0 & -0.2 & 0.2 \\
 & Supply-driven PCE Inflation & 1969M12 & 2024M4 & 653 & 1.8 & 1.5 & -0.8 & 6.2 \\
 & Demand-driven PCE Inflation & 1969M12 & 2024M4 & 653 & 1.1 & 0.8 & -0.8 & 4.0 \\ \hline
\multirow{5}{*}{Controls} & Headline PCE Inflation Rates & 1960M1 & 2024M4 & 772 & 3.2 & 2.3 & -1.5 & 11.0 \\
 & Adjusted NFCI & 1971M1 & 2024M4 & 640 & 0.0 & 1.0 & -1.4 & 5.3 \\
 & NFCI - Nonfinancial leverage   subindex & 1971M1 & 2024M4 & 640 & 0.0 & 1.0 & -2.0 & 2.7 \\
 & Term Spread (10 Year-2 Year) & 1976M6 & 2024M4 & 575 & 0.9 & 0.9 & -2.1 & 2.8 \\ \hline
\multirow{1}{*}{Robustness} & Wage Inflation Rates & 1965M1 & 2024M4 & 712 & 4.1 & 1.8 & 1.2 & 8.9 \\
 \hline\hline
\end{tabular}}  
\end{center}
\label{table:Summary_Statistics}
{\footnotesize \emph{\bf Notes to Table}: This table shows the first observation (First), last observation (Last), number of observations (Obs), mean, standard deviations (SD), minimum (Min), and maximum (Max) for the dependent variables, instruments, and controls. The units are percentage points.} \\ 
{\footnotesize \emph{\bf Source}: Authors' calculation.} \\ 
\end{table}

Table \ref{table:Summary_Statistics} reports the summary statistics and sample period of data considered in the model.

\subsection{Why IVQR? Why not linear IV regression?}\label{sec_2sls_ivqr}

A linear IV regression assumes the same relationship among the dependent variable, the regressor of interest, and its instrument across the quantiles of the dependent variable. In our case, these are changes in the unemployment rate, inflation, and the structural driver of inflation, respectively. However, if the relationship among these variables systematically varies across the quantiles---particularly in the upper quantiles, such as during economic recessions---the linear IV model will fail to capture these nonlinear dynamics. As a result, the linear IV model is likely to perform poorly in predicting changes in the unemployment rate during recessions.


\begin{figure}[t]
\caption{IV Linear Regression v. IV Quantile Regression}\label{figure:why_IVQR}
	\begin{center}
     \begin{subfigure}[b]{0.65\textwidth}
         \centering
         \textbf{Panel A}. Linear IV (Supply-driven) 
         \includegraphics[width=1\textwidth, height = 0.5\textwidth]{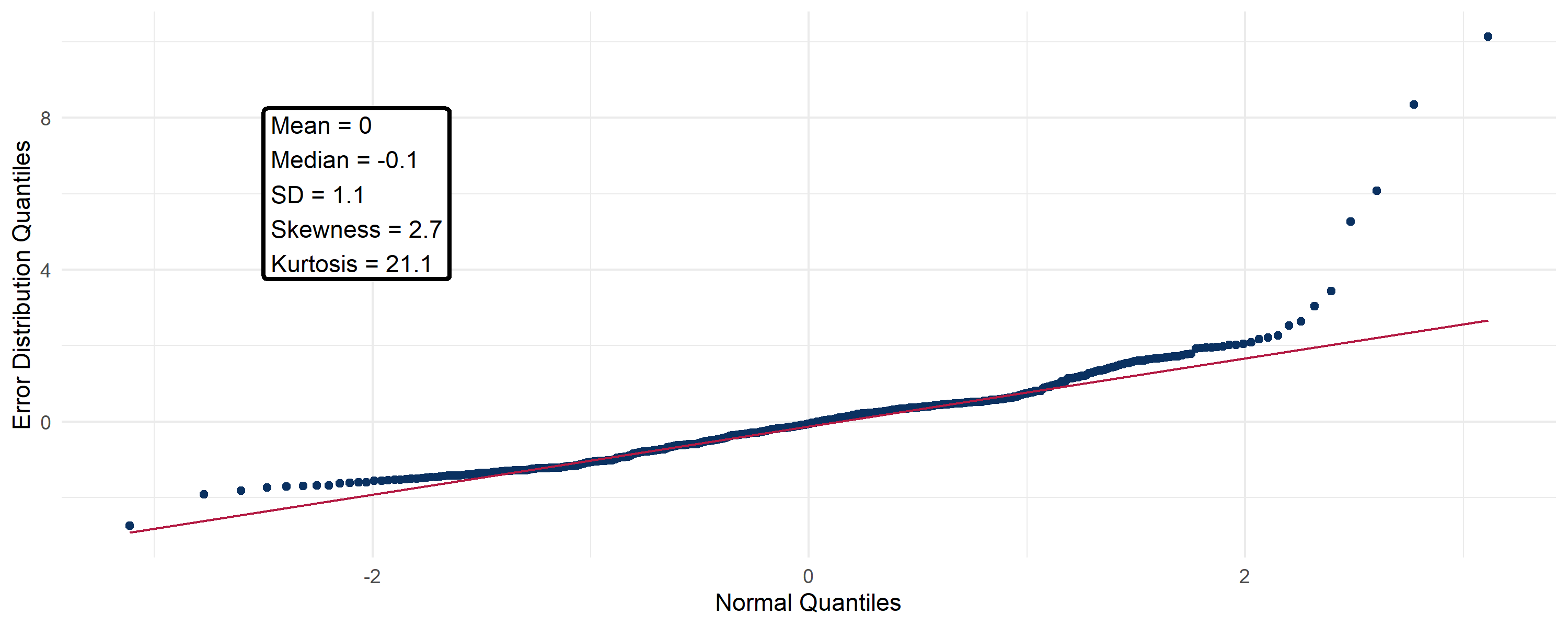}
     \end{subfigure}      
     \medskip 
	\begin{subfigure}[b]{0.49\textwidth}
         \centering
         \textbf{Panel B}. IVQR (Supply-driven)
         \includegraphics[width=1\textwidth, height = 0.6\textwidth]{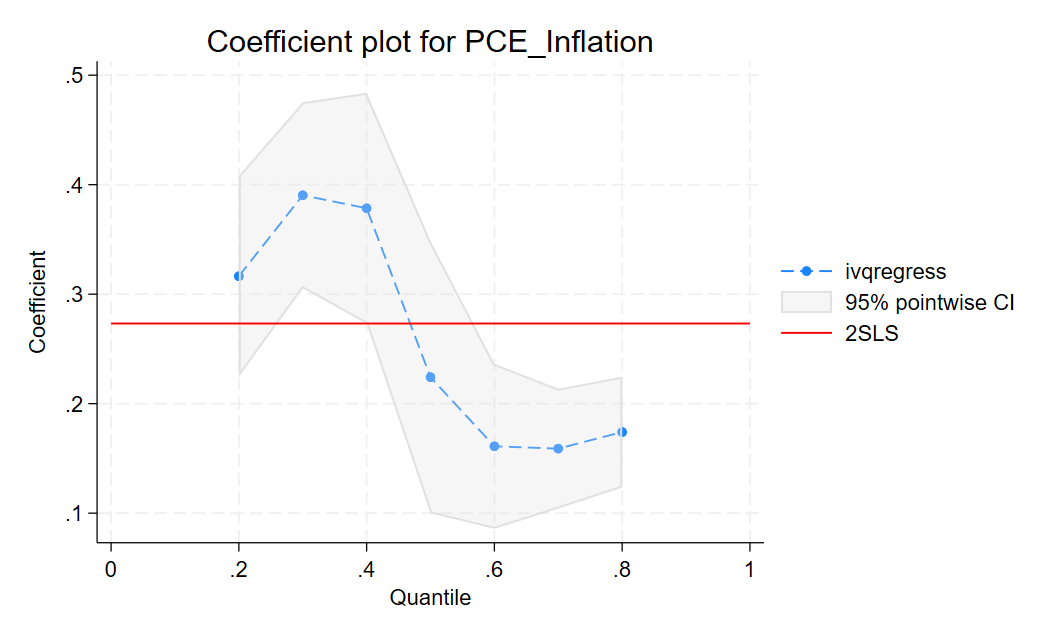}
     \end{subfigure}
     \begin{subfigure}[b]{0.49\textwidth}
         \centering
         \textbf{Panel C}. IVQR (Demand-driven)
         \includegraphics[width=1\textwidth, height = 0.6\textwidth]{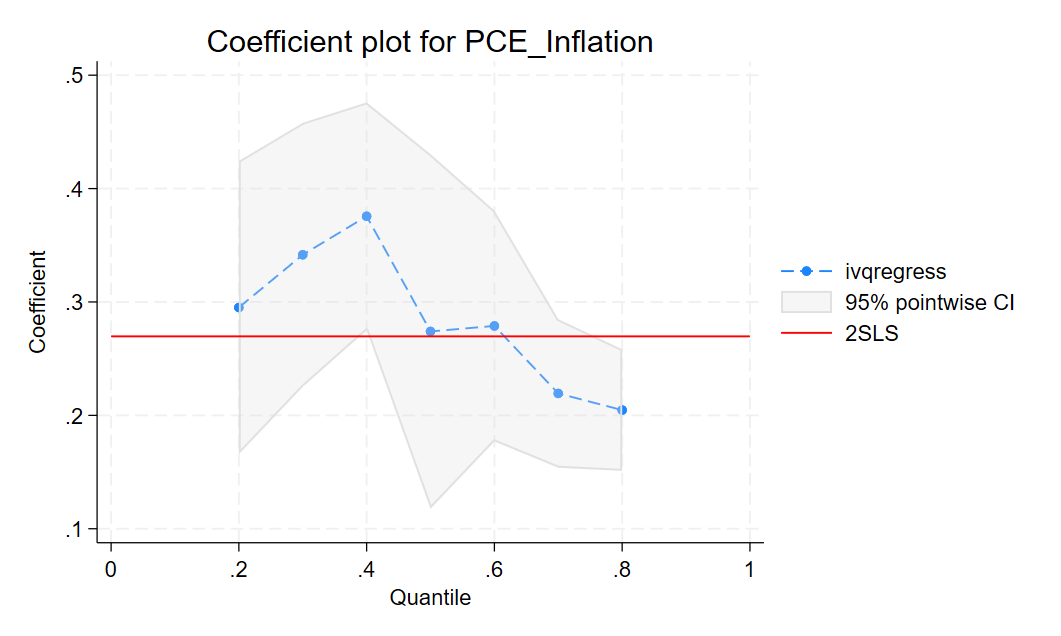}
     \end{subfigure}      
     \end{center} 
{\footnotesize \textbf{Notes to figure}. Panel A shows the Quantile-Quantile plot (QQ plot) of the estimation errors from the linear instrumental variable regressions against the normal distribution. Panels B and C show the coefficients estimated from the IVQR across multiple quantiles. The dependent variables are the three-year change of the unemployment rates for Whites. And the instruments are the year-over-year change of the supply-driven and demand-driven headline inflation constructed by \cite{shapiro2022much}. \textbf{Source:} Authors' calculation.} 
\end{figure}

Figure \ref{figure:why_IVQR} illustrates this point. To answer this question, we first estimate both the linear IV and IVQR models for White workers. The IVQR model is the same as the one used in our empirical analysis, which is detailed in the next section. In this model, the dependent variable is the three-year change in unemployment rates of White workers, the regressor is PCE price inflation, and the instruments are supply- and demand-driven inflation measures constructed by \cite{shapiro2022much}.

Panel A displays the error terms of the linear IV model across quantiles (dark blue dots, Y-axis) against a normal distribution (X-axis). The red line depicts the quantile distribution that would be expected if the error terms were normally distributed. As shown in the panel, the error terms of the linear IV model deviate substantially from normality, particularly in the lower and upper quantiles. The inset box in Panel A shows that the error distributions exhibit heavy right tails, as evidenced by positive skewness and excess kurtosis. These leptokurtic errors suggest that the linear model performs poorly in predicting changes in unemployment at the tails of the distribution.

Panels B and C present the coefficients estimated from the IVQR and 2SLS models for White workers. While the 2SLS estimates are very similar between supply-driven and demand-driven inflation, the IVQR results reveal substantial heterogeneity across quantiles. Thus, relying solely on 2SLS estimates risks overlooking important differences in the effects of demand- and supply-driven inflation at the distribution’s tails.

All told, the linear IV regression model may fail to capture the nuanced effects of supply- and demand-driven inflation on unemployment tail risks, making the IVQR a more suitable empirical approach for our research question.

\section{Empirical Results}\label{sec:results}

Section \ref{sec:aggregate_result} presents the estimation results for the aggregate case. Section \ref{sec:disaggregate_result} reports the results for disaggregated groups. 

\subsection{Aggregate Unemployment Risk}\label{sec:aggregate_result}

\begin{table}[t]
    \caption{Quantile Regressions and IV Quantile Regressions}  
    \begin{center}
    \scalebox{0.9}{
    \begin{tabular}{l|c|c|c|c|c|c}
\hline \hline 
\multicolumn{1}{l|}{\textbf{}} & \multicolumn{3}{c|}{\textbf{1-year change}} & \multicolumn{3}{c}{\textbf{3-year change}} \\ \hline
\multicolumn{1}{l|}{\textbf{}} & \multicolumn{1}{c|}{\textbf{{[}1{]} QR}} & \multicolumn{1}{c|}{\textbf{\begin{tabular}[c]{@{}c@{}}{[}2{]} IVQR \\ Supply\end{tabular}}} & \multicolumn{1}{c|}{\textbf{\begin{tabular}[c]{@{}c@{}}{[}3{]} IVQR \\ Demand\end{tabular}}} & \multicolumn{1}{c|}{\textbf{{[}4{]} QR}} & \multicolumn{1}{c|}{\textbf{\begin{tabular}[c]{@{}c@{}}{[}5{]} IVQR \\  Supply\end{tabular}}} & \multicolumn{1}{c}{\textbf{\begin{tabular}[c]{@{}c@{}}{[}6{]} IVQR \\ Demand\end{tabular}}} \\ \hline
 \\ \hline
\textbf{PCE Inflation} & 0.12*** & 0.23*** & 0.01 & 0.20*** & 0.19*** & 0.25*** \\
\textbf{} & (0.04) & (0.04) & (0.06) & (0.03) & (0.03) & (0.03) \\
\textbf{Unemployment Rate} & -0.28*** & -0.36*** & -0.20** & -0.37*** & -0.36*** & -0.43*** \\
\textbf{} & (0.08) & (0.08) & (0.10) & (0.06) & (0.06) & (0.07) \\
\textbf{ANFCI} & 0.78*** & 0.75*** & 0.78*** & -0.30*** & -0.29*** & -0.27*** \\
\textbf{} & (0.10) & (0.11) & (0.14) & (0.07) & (0.06) & (0.07) \\
\textbf{NFCI\_NFL} & 0.30*** & 0.31*** & 0.33*** & 1.58*** & 1.58*** & 1.53*** \\
\textbf{} & (0.10) & (0.08) & (0.10) & (0.07) & (0.05) & (0.05) \\
\textbf{Term Spread} & 0.38*** & 0.57*** & 0.18 & -0.05 & -0.08 & 0.02 \\
\textbf{} & (0.11) & (0.11) & (0.18) & (0.08) & (0.12) & (0.12) \\
\textbf{Constant} & 1.45*** & 1.50*** & 1.51*** & 2.15*** & 2.12*** & 2.28*** \\
\textbf{} & (0.42) & (0.39) & (0.40) & (0.29) & (0.24) & (0.26) \\ \hline
    \end{tabular}}
    \end{center}
    \label{table:Linear_vs_Quantile_2}
    {\footnotesize \emph{\bf Notes to table:} This table shows the estimates of the quantile regression at the 80\% quantile over the period 1976M6 to 2021M6. QR is the regular quantile regression, and IVQR-Supply Shock and IVQR-Demand Shock are the inverse quantile estimates with the same instruments. The numbers in parentheses are the standard errors. The *, **, and *** denote statistical significance at the 10 percent, 5 percent, and 1 percent levels, respectively. \emph{\bf Source:} Authors' calculation.}
\end{table}

This section focuses on the unemployment risks at the 80\% quantile, highlighting the need to distinguish inflation effects across different time horizons and their structural sources. Table \ref{table:Linear_vs_Quantile_2} presents the coefficient estimates from the quantile regression model, comparing the results with and without using supply- and demand-driven inflation as instrumental variables. We focus on two time horizons: one year and three years ahead. The former captures the timeframe relevant for recession prediction, while the latter reflects the medium-term horizon over which monetary policy effects unfold throughout the economy.

Inflation has varying effects on the unemployment tail risk at one-year and three-year horizons. Notably, it has a greater effect on unemployment tail risk three years ahead than one year ahead (Columns [1] and [4]). When distinguishing inflation by its structural sources, supply-driven inflation significantly raises unemployment risk one year ahead (Column [2]), while the effect of demand-driven inflation is negligible (Column [3]). However, three years ahead, both supply-driven and demand-driven inflation increase unemployment risk, with demand-driven inflation demonstrating stronger predictive power (Columns [5] and [6]). In summary, supply-driven inflation is a key predictor of short-term unemployment risk, whereas both types of inflation predict medium-term unemployment risks. This finding underscores the importance of considering both supply- and demand-driven inflation, as well as the varying time horizons over which inflationary pass-through affects unemployment tail risks.

Financial indicators also exhibit differential effects between the two forecasting horizons. The ANFCI \textemdash the indicator of general financial condition \textemdash raises the risk one year ahead (Column [1]) but lowers the risk three years ahead (Column [4]). In addition, the term spread raises the one-year-ahead risk, while its three-year-ahead effect is not statistically significant. The index of credit cycle \textemdash NFCI-NFL \textemdash has a larger statistically significant effect in predicting the unemployment tail risk three years ahead than one year ahead. In summary, the general financial condition and the term spread raise the short-run unemployment risk, while the credit cycle raises the long-run risk. \textcolor{black}{These findings align with previous literature, which suggests that credit quantities influence the economy at a lower frequency than credit prices (\citealp{hamilton2020measuring}). From an economic perspective, during a credit expansion, the economy typically experiences growth, characterized by low unemployment and loose financial conditions. However, this period of expansion also increases vulnerabilities, such as higher leverage on the balance sheets of households and firms. When economic shocks occur, the economy begins to deleverage, and financial distress manifests primarily through the widening of credit spreads (\citealp{adrian2022term}, \citealp{plagborg2020growth}).}

\subsection{Who's at Risk? Disaggregate Unemployment Risk} \label{sec:disaggregate_result}

Next, we examine cross-sectional heterogeneity in unemployment tail risks, with a particular focus on race and reason for unemployment. While we consider various other worker attributes, we highlight these two categories because they exhibit the most pronounced heterogeneity. 

\subsubsection{Summary of Disaggregate Results} 

Table \ref{tab:summary} summarizes the distributional effects of the two types of inflation on the unemployment tail risks of various groups.\footnote{The complete set of estimates is reported in Section B of the online appendix. Section G provides the numerical estimates and their standard errors, while Section F reports the estimates across all quantiles.}

Two key observations emerge. \textcolor{black}{First, among the various worker attributes considered—including demographic characteristics, education, job status, and reasons for unemployment—cross-sectional heterogeneity is most pronounced by race and by reason for unemployment. Second, the supply-driven inflation has more pervasive and immediate distributional effects. Unlike the supply-driven inflation, the demand-driven inflation does not show statistically significant differential effects in the short term. In the medium term, the supply-driven inflation affects all categories considered, disproportionately raising unemployment tail risks for cyclically vulnerable groups---such as racial and ethnic minorities, the less-educated, part-time workers, job losers, and young workers. In contrast, the demand-driven inflation widens disparities in the medium term with statistical significance, limited to race and reason for unemployment. Nonetheless, it is important to note that demand-driven inflation significantly raises the unemployment tail risk for Black workers compared to White workers, and lowers the risk for job leavers while raising it for job losers---a pattern and magnitude that are similar to those observed under supply-driven inflation. Despite reflecting strong demand, the persistent feature of demand-driven inflation raises unemployment tail risks, with disproportionately adverse effects on racial and ethnic minorities and job losers, while also discouraging job quits.} 

All told, supply-driven inflation has a more immediate impact on exacerbating inequality in the short run, while both supply- and demand-driven inflation contribute to distributional effects over the longer run. 

\begin{table}[t]
    \caption{Summary of Who-at-Risk}
    \centering
    \begin{tabular}{l|c|c|c|c}
    \hline \hline 
      & \multicolumn{2}{c|}{\bf 1-year ahead} & \multicolumn{2}{c}{\bf 3-year ahead} \\ \hline
                  & Supply   & Demand   & Supply   & Demand   \\ \hline 
    Race (Blacks)         & \checkmark   &      & \checkmark   & \checkmark  \\ \hline
    Education (less-educated)    &              &      & \checkmark   &    \\ \hline
    Age/gender (young)   &              &      & \checkmark  &   \\ \hline
    Reason (job losers)   & \checkmark   &      & \checkmark  & \checkmark \\ \hline
    Job status (part time) &              &      & \checkmark  &    \\ \hline \hline 
    \end{tabular} \\ 
    \medskip 
    \raggedright 
{\footnotesize \emph{\bf Notes to Table}: In this table, 
$\checkmark$ indicates the type of inflation and the category of worker attributes that shows statistically significant differential effects in the cross section. The parentheses of the first column show the worker attribute with the highest overall unemployment tail risks.} \label{tab:summary}
\end{table}

\begin{figure}[htbp]
\caption{Unemployment Tail Risks by Reason for Unemployment}
\footnotesize
\medskip
    \centering
    Panel A. Demand-driven (1-year) \\ 
    \includegraphics[width=0.7\textwidth, height = 0.25\textwidth]{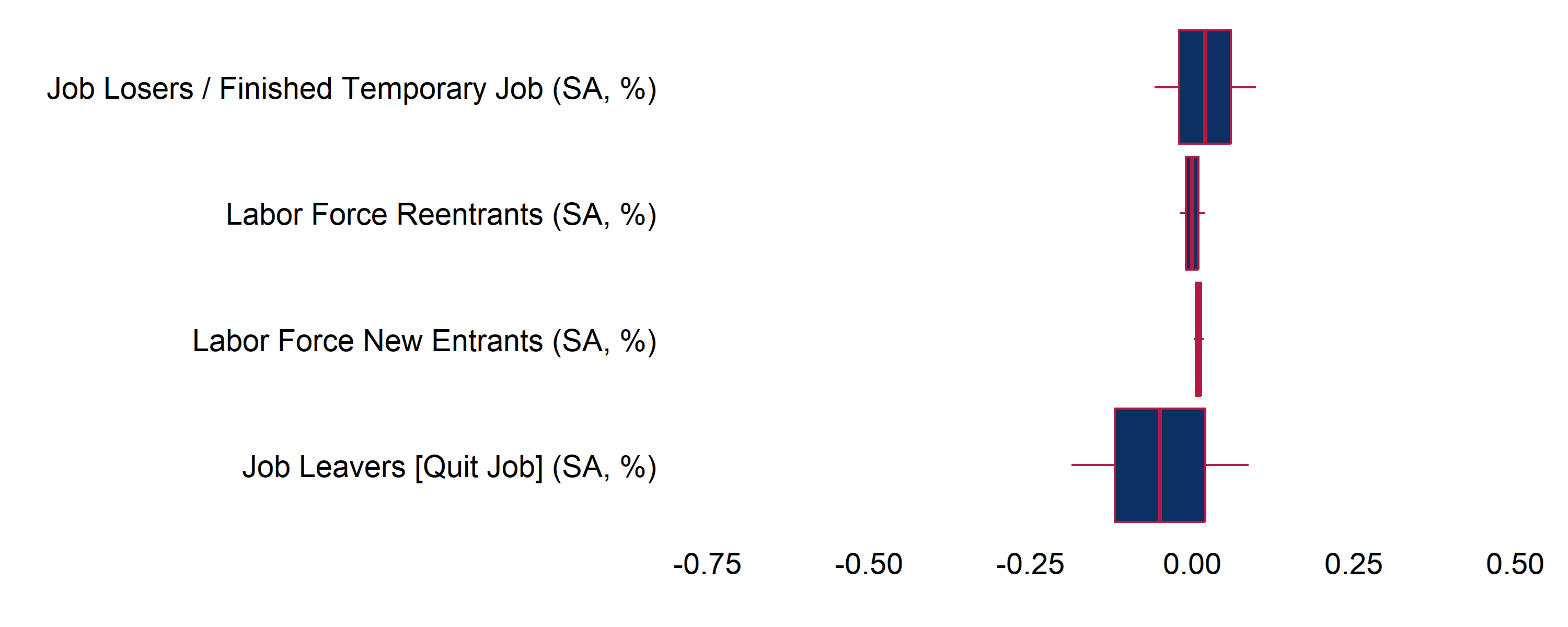} \\
    Panel B. Supply-driven (1-year) \\ 
    \includegraphics[width=0.7\textwidth, height = 0.25\textwidth]{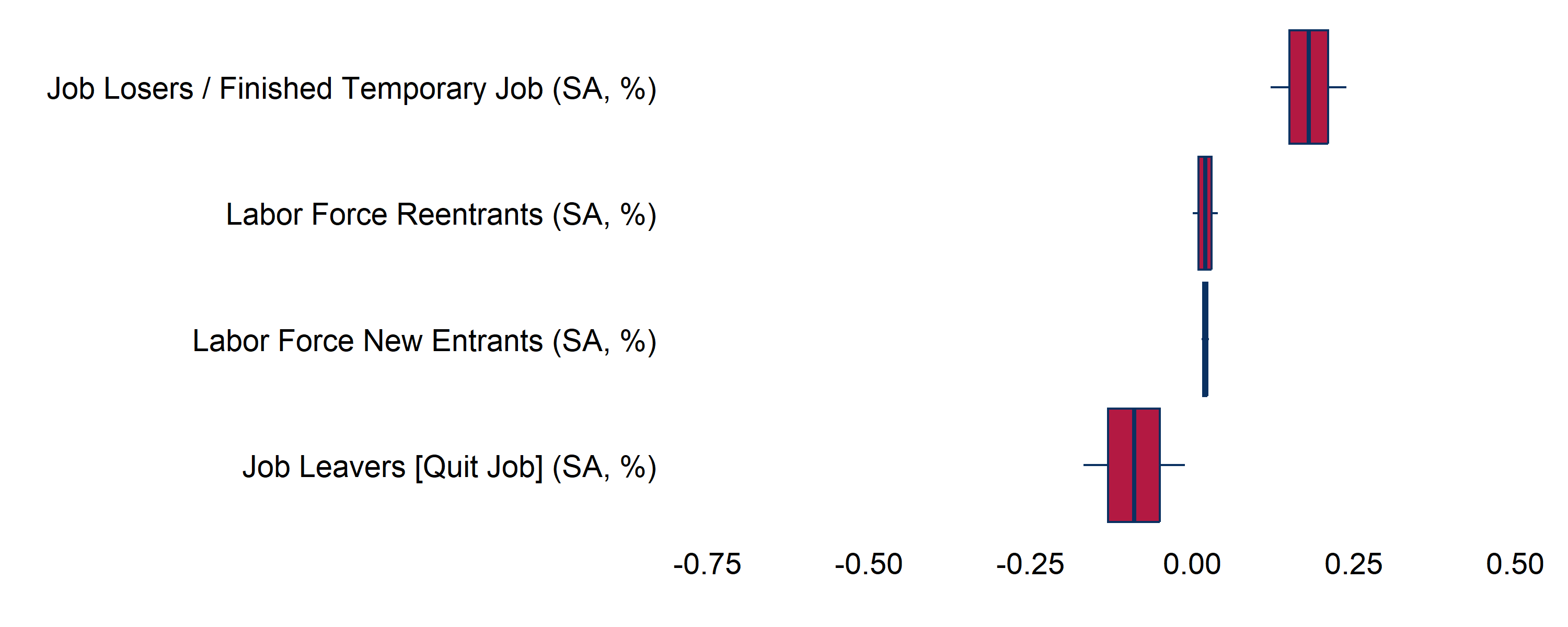} \\ 
    Panel C. Demand-driven (3-year) \\ 
    \includegraphics[width=0.7\textwidth, height = 0.25\textwidth]{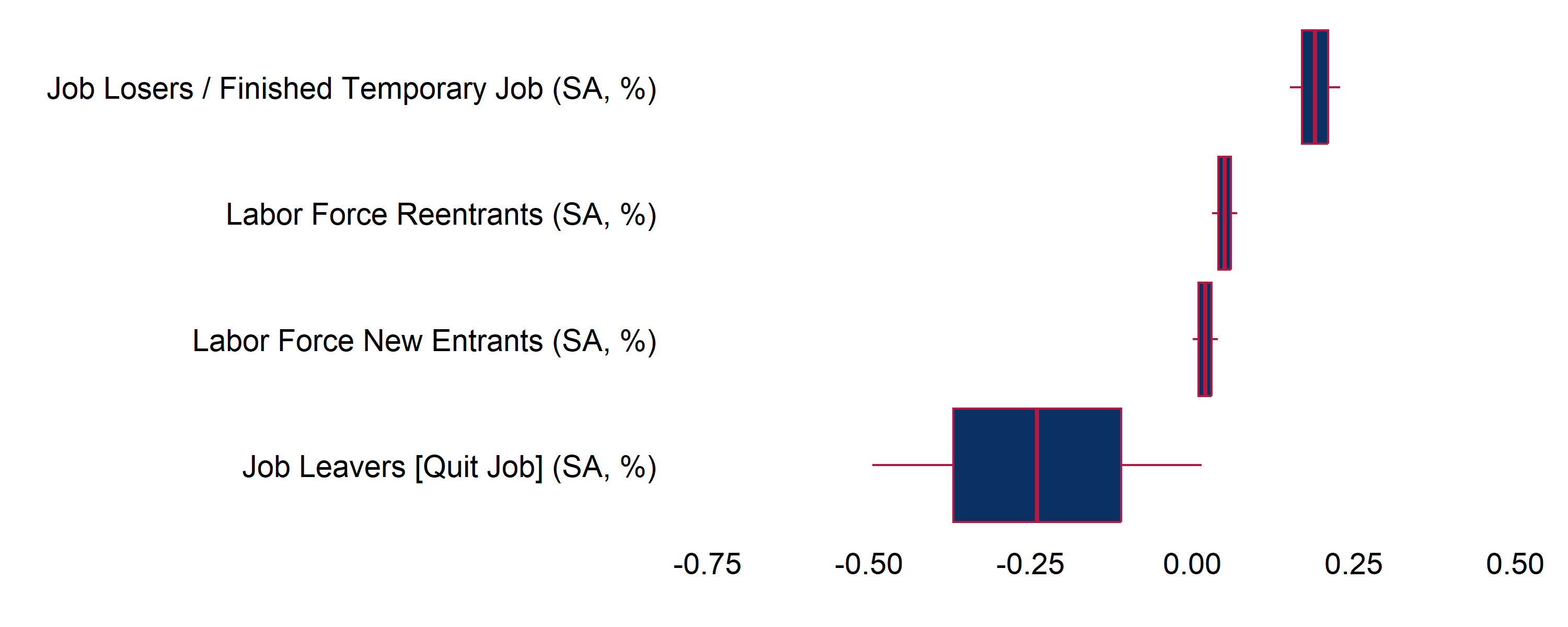} \\
    Panel D. Supply-driven (3-year) \\ 
    \includegraphics[width=0.7\textwidth, height = 0.25\textwidth]{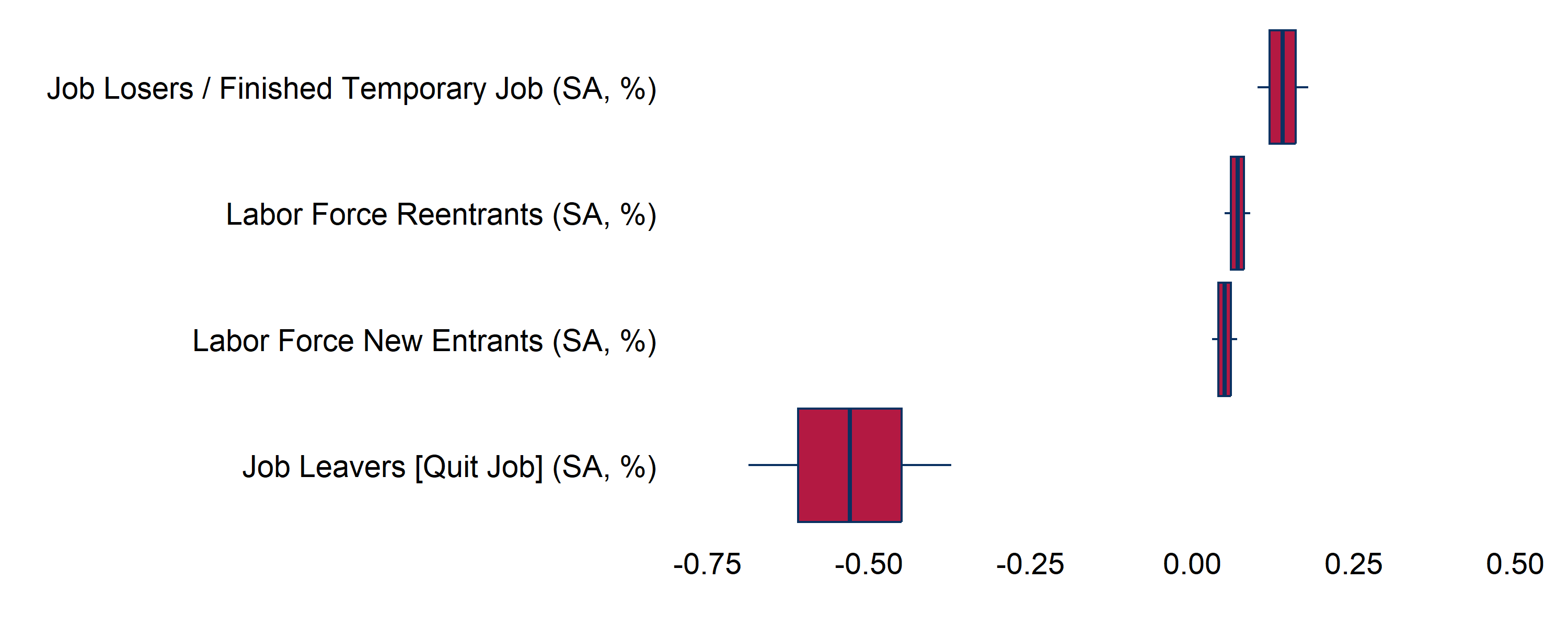} \\ 
    \raggedright 
{\footnotesize \textbf{Notes to figure:} This figure shows the response of 1-year and 3-year changes in unemployment rates to a one-percentage-point increase in PCE price inflation, estimated using quantile regressions at the 80th percentile. Panels A and C use the demand-driven inflation as the instrument, while Panels B and D use the supply-driven inflation, both from \cite{shapiro2022much}. The box plots depict variations within one and two standard deviations. \textbf{Source:} Authors' calculation.}
    \label{fig:reason1}
\end{figure}

\subsubsection{Reason for Unemployment}

Figure \ref{fig:reason1} shows the inflation coefficients by source and forecast horizon for workers unemployed for different reasons---job loss, labor force re-entry, new entry into the labor force, and quits.\footnote{We consider those whose temporary jobs ended as job losers (involuntary separation).} It is notable that supply-driven inflation lowers the unemployment risks of job quitters while raising those of the others for one-year and three-year ahead (Panels B and D), and so does the demand-driven inflation for three-year ahead (Panel C). Inflation raises the unemployment risks of job losers the most, followed by those of re-entrants to the labor force and new entrants. 

Specifically, a one percentage-point increase in supply-driven inflation raises the one-year-ahead unemployment rate of job losers by 0.2 percentage point, increases that of reentrants and new entrants by about 0.05 percentage point, but lowers the unemployment rate of job leavers by 0.1 percentage point (Panel B). The difference in coefficients between job losers and job leavers is statistically significant. Notably, the negative effects of inflation on the unemployment rate of job leavers are even larger in the medium term: a one percentage-point increase in demand-driven inflation reduces their unemployment rate by 0.25 percentage point (Panel C), while an equivalent increase in supply-driven inflation lowers it by 0.5 percentage point (Panel D). Meanwhile, the upward effect on the unemployment rate of job losers remains relatively consistent, with coefficients ranging from approximately 0.2 to 0.25 percentage points (Panels B-D). 


The observed differences in unemployment risks between job leavers and job losers can be explained as follows. Facing increased production costs, whether from higher input prices or increased wages driven by labor demand, workers are less likely to quit their jobs and more likely to accept new job offers when searching for a new job. This lowers the unemployment rate for job leavers. Facing inflation, regardless of the source of inflation, workers become less likely to quit their jobs and take the risk of  unemployment following job quits. Rising prices raise concerns about reducing consumption, which discourages workers from moving to jobs with higher productivity and makes them less willing to take the risk of unemployment.

Meanwhile, the increased cost pressure likely induces firms to lay off low-productivity workers and reduce demand for such workers, raising the unemployment risk for job losers. This illustrates that the same structural inflationary shocks can have different effects on voluntary and involuntary separations. \footnote {\cite{AV2023} also emphasize that heterogeneity in separation is key to understanding the distributional effects in the labor market of cyclical shocks and monetary policy.} The stark heterogeneity in unemployment risk based on reasons for unemployment is consistent with findings regarding the importance of reasons for unemployment in unemployment hazards (e.g., \citealp{HS2018}; \citealp{ahn2023}).

All told, it is important to note that both types of inflation raise the unemployment risks of job losers in the medium term, suggesting that higher inflation---regardless of its source---raises job destruction or involuntary separation. The link between inflation and unemployment risks suggests the importance of considering the effect of inflation on involuntary separation and voluntary separation in the labor market and recession predictions. 

\subsubsection{Race}

\begin{figure}
\caption{Unemployment Tail Risks by Race}
\medskip \footnotesize 
    \centering
    Panel A. Demand-driven (1-year) \\ 
    \includegraphics[width=0.7\textwidth, height = 0.25\textwidth]{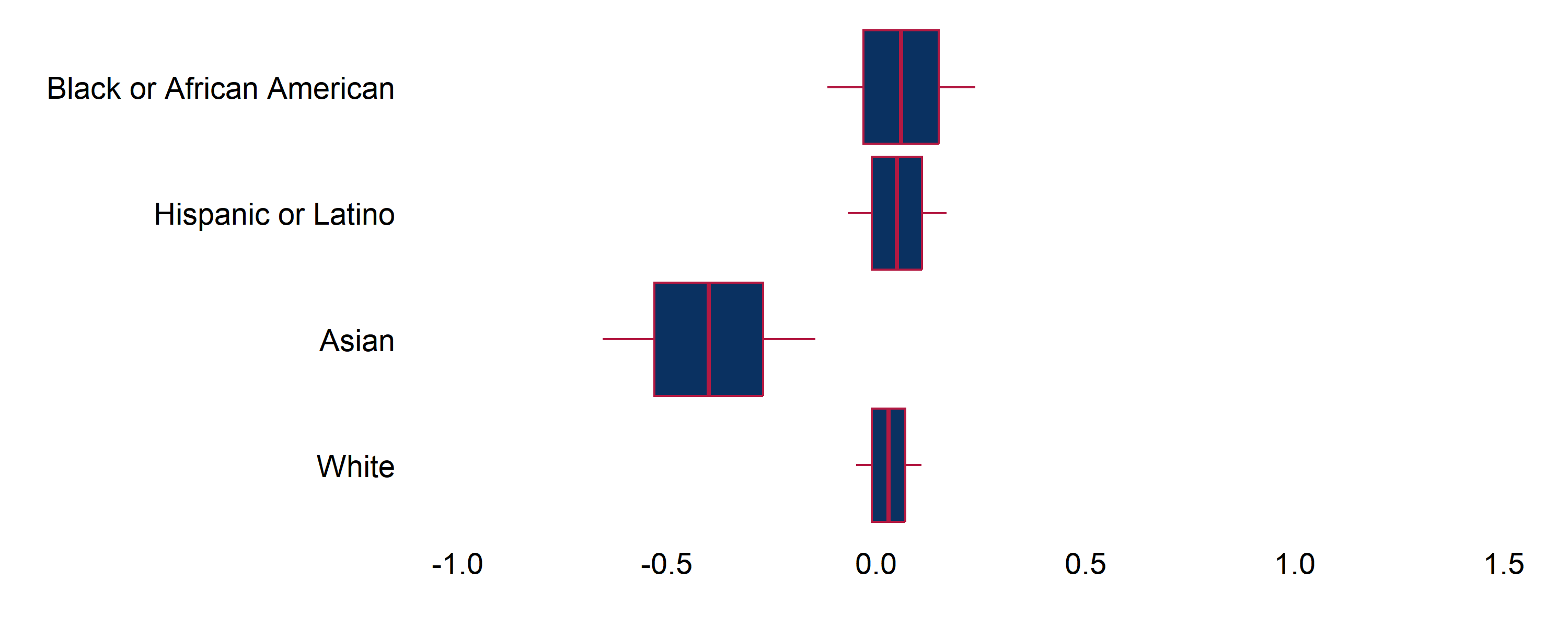} \\
    Panel B. Supply-driven (1-year) \\ 
    \includegraphics[width=0.7\textwidth, height = 0.25\textwidth]{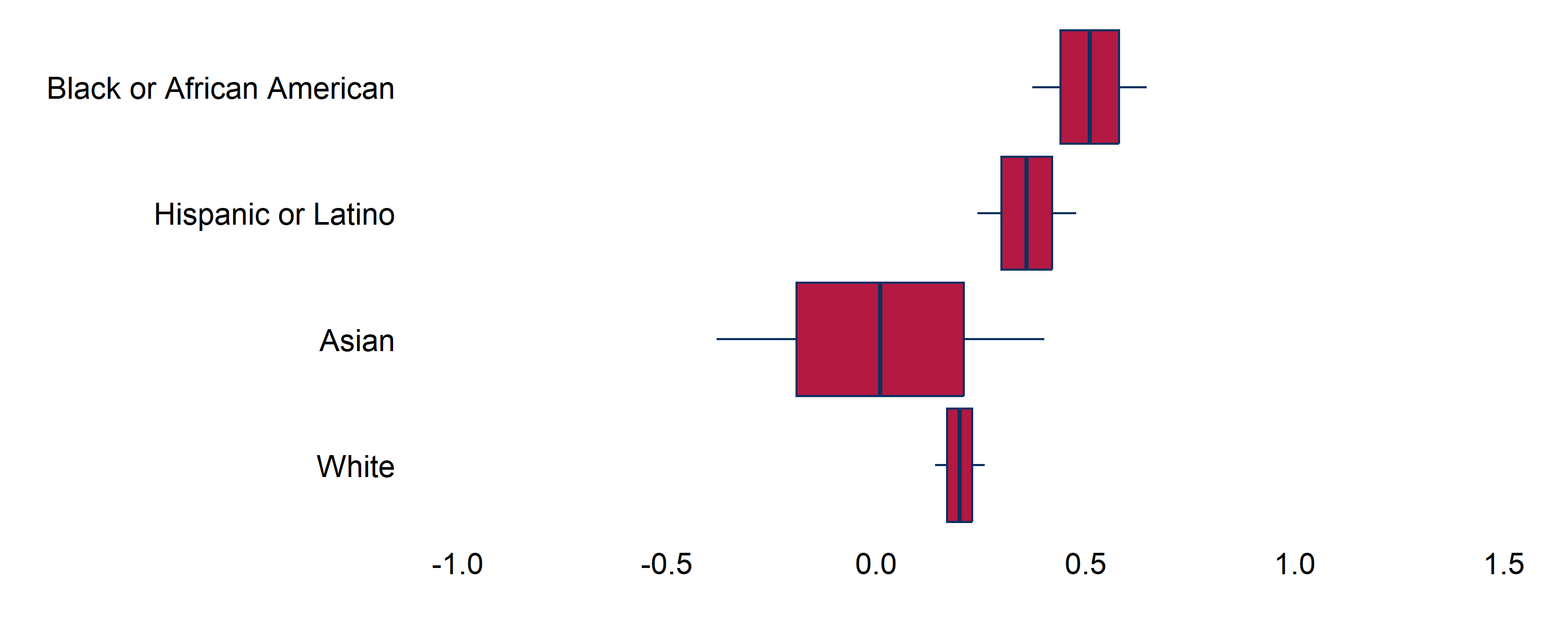} \\ 
    Panel C. Demand-driven (3-year) \\ 
    \includegraphics[width=0.7\textwidth, height = 0.25\textwidth]{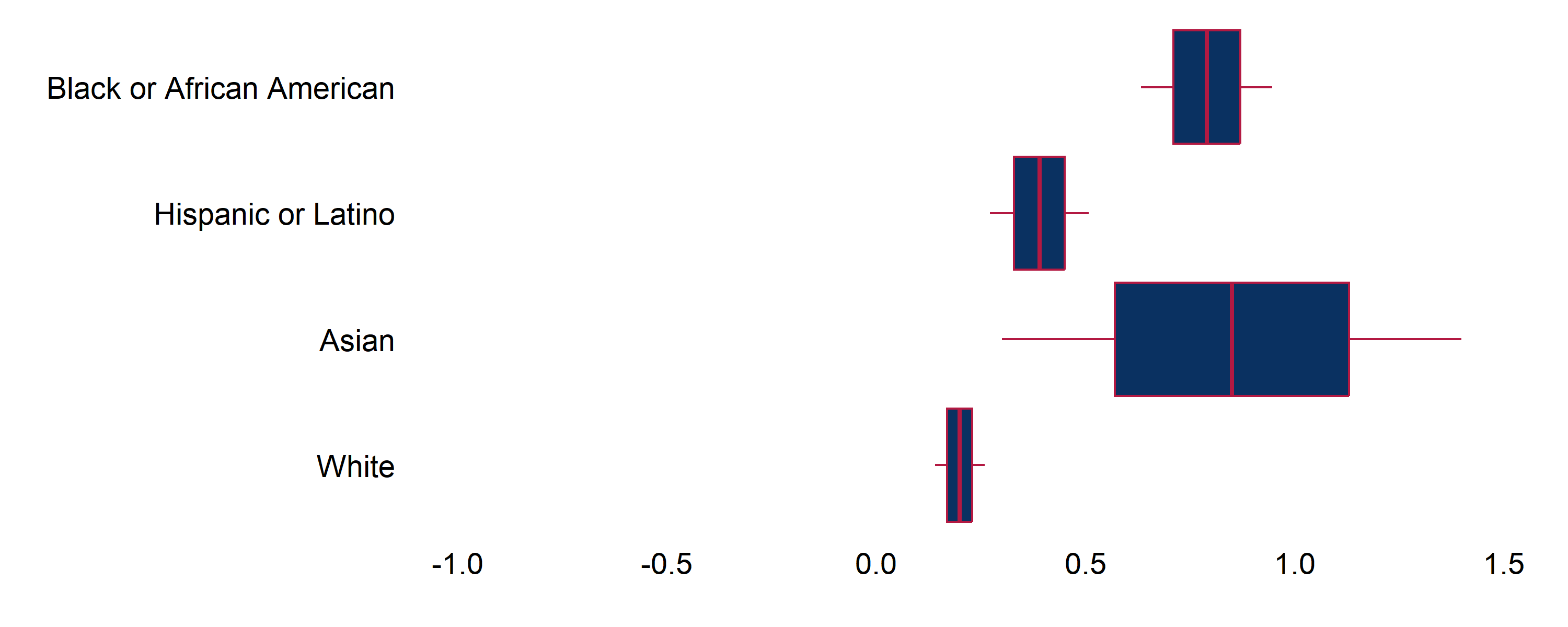} \\
    Panel D. Supply-driven (3-year) \\ 
    \includegraphics[width=0.7\textwidth, height = 0.25\textwidth]{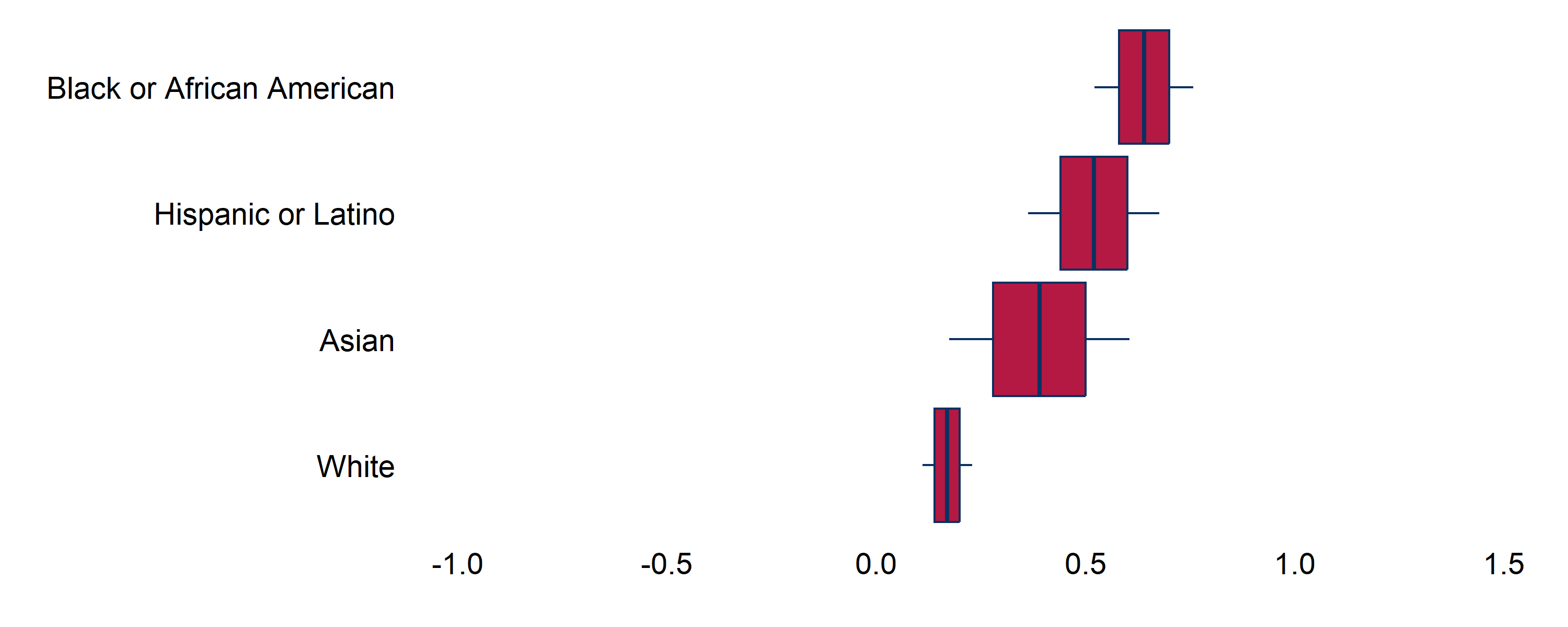} \\ 
\raggedright 
{\footnotesize \textbf{Notes to figure:} This figure shows the response of 1-year and 3-year changes in unemployment rates to a one-percentage-point increase in PCE price inflation, estimated using quantile regressions at the 80th percentile. Panels A and C use the demand-driven inflation as the instrument, while Panels B and D use the supply-driven inflation, both from \cite{shapiro2022much}. The box plots depict variations within one and two standard deviations. \textbf{Source:} Authors' calculation.}
    \label{fig:race}
\end{figure}

Figure \ref{fig:race} displays the effects of demand-driven and supply-driven inflation on the unemployment risks of racial groups. Overall, supply-driven inflation raises the unemployment risks of non-White workers \textemdash Blacks the most, followed by Hispanics and Latinos \textemdash significantly more than those of White workers in both the one-year-ahead and three-year-ahead horizons. Specifically, supply-driven inflation raises the unemployment rate of Black workers by about 0.5 percentage point one year ahead and 0.6 percentage point three years ahead, compared to just 0.2 percentage point for White workers in both the short and medium term. 

Demand-driven inflation has also differential effects across racial groups, but only in the medium run---raising tail risk the most for Black workers, followed by Hispanic and Latino workers, with the smallest upside effect observed for White workers.\footnote{The effects on Asians are imprecisely estimated owing to the small sample size, as shown by the large confidence intervals, which make the between-group comparison statistically insignificant.} Specifically, a one percentage-point increase in demand-driven inflation raises the unemployment rate of Black workers by more than the effect of supply-driven inflation. 
It is important to note that the differential effects of demand-driven inflation between Black and White workers are greater than those of supply-driven inflation. In summary, supply-driven inflation has more immediate distributional effects across racial groups compared to demand-driven inflation, although both types of inflation exhibit distributional effects in the medium term.\footnote{This result aligns with the findings of \cite{cajner2017racial}, which show that Black workers' joblessness is more cyclically sensitive than that of White workers, consistent with the interpretation of supply-driven inflation as a recessionary shock.}

Interestingly, unemployment tail risks by reason for unemployment and by race exhibit similar patterns. Supply-driven inflation exhibits differential impacts between job leavers and job losers, as well as between Black and White workers, for both one-year and three-year horizons. Also, demand-driven inflation shows such heterogeneity for three-year-ahead unemployment tail risks. This association suggests that race and reason for unemployment is related. Indeed, White workers are more likely to be voluntary job leavers than Black workers. Specifically, job leavers accounted for 11 percent of White workers but only 7 percent of Black workers in 1994.\footnote{This gap narrowed over time, reaching 14 percent for Whites and 13 percent for Blacks in 2019.} However, reason for unemployment is not the necessary or sufficient condition for racial disparities. To illustrate, the share of job losers is greater among Whites than Blacks, while the share of re-entrants to the labor force is higher among Blacks than Whites. The combined share of job losers and re-entrants is similar between the two groups. This suggests that while differences in the propensity to quit help explain the Black-White gap in unemployment tail risks, reason for unemployment alone does not account for racial disparities. 


\section{Structural Drivers: Oil, Monetary Policy and Wage Setting} \label{sec:robustness}

We examine the extent to which supply- and demand-driven inflation are associated with structural shocks, focusing in particular on oil shocks and monetary policy shocks. Finally, we compare the effects of price inflation and wage inflation to assess the role of wage setting in shaping unemployment tail risks. 

\subsection{Oil Supply Shock} \label{sec:oil} 


We employ the oil supply shock as the external instrument of supply-side driver of inflation, and examine the effects of oil shocks on the unemployment tail risks. An important harbinger of both headline inflation and economic recessions is a large increase in oil prices (\citealp{hamilton2009causes}). \cite{del2023inflationary} also found that oil price increases have more adverse effects on disadvantaged households whose budgets depend on the price of motor fuel. 

\begin{figure}[h!]
	\centering
 \caption{Effects of Oil Supply Shock on Unemployment Risks} \footnotesize
 \bigskip
\begin{subfigure}[b]{0.49\textwidth}
         \centering \footnotesize
         Panel A. Oil-Supply : Race (1 year) 
         \includegraphics[width=\textwidth, height=0.5\textwidth]{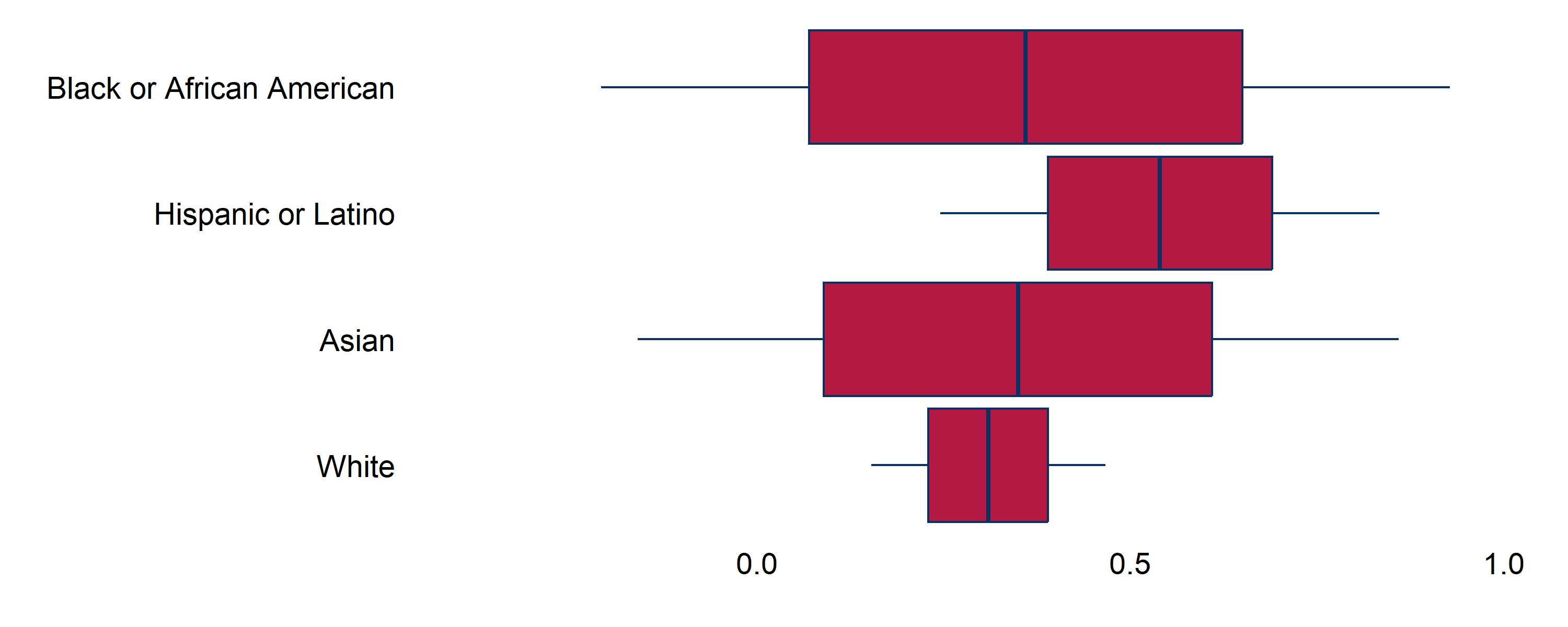}
     \end{subfigure}
     \hfill
     \begin{subfigure}[b]{0.49\textwidth}
         \centering \footnotesize
         Panel B. Oil-Supply : Reason (1 year)
         \includegraphics[width=\textwidth, height=0.5\textwidth]{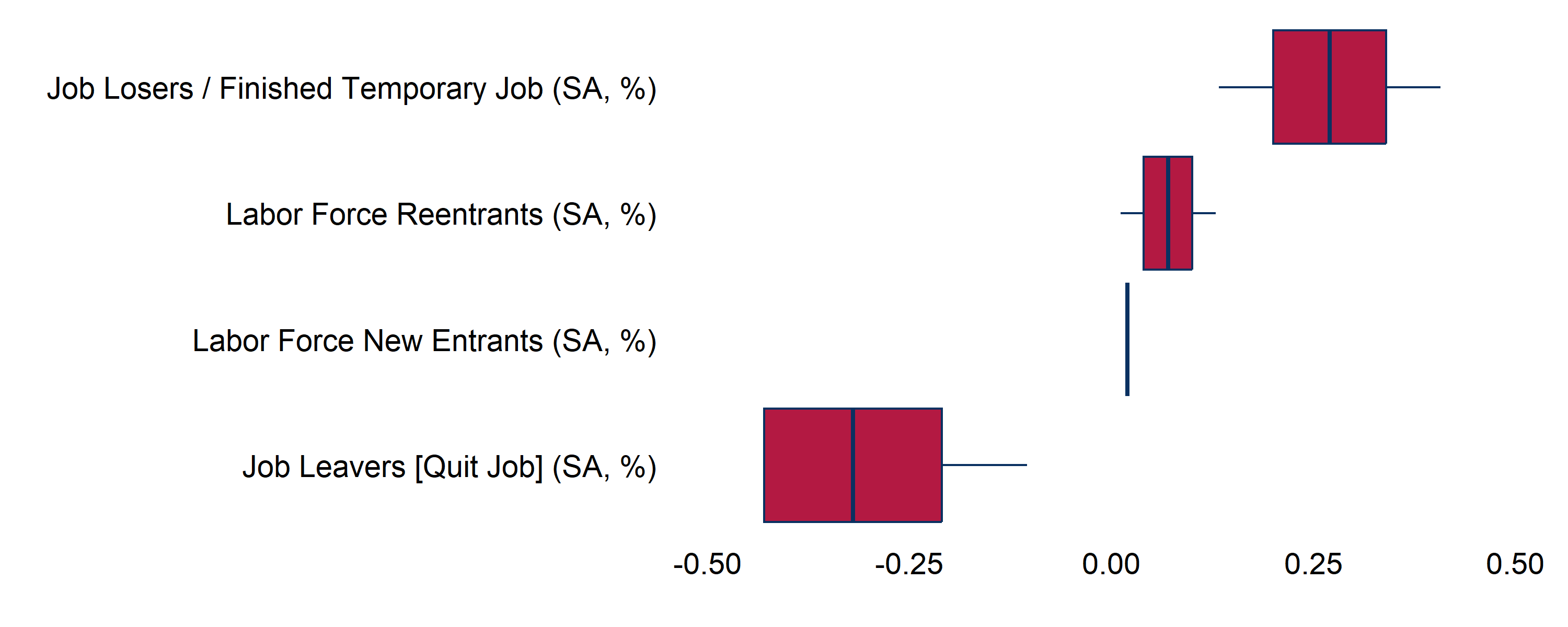}         
     \end{subfigure}
\medskip 
\begin{subfigure}[b]{0.49\textwidth}
         \centering \footnotesize
         Panel C. Overall Supply : Race (1 year) 
         \includegraphics[width=\textwidth, height=0.5\textwidth]{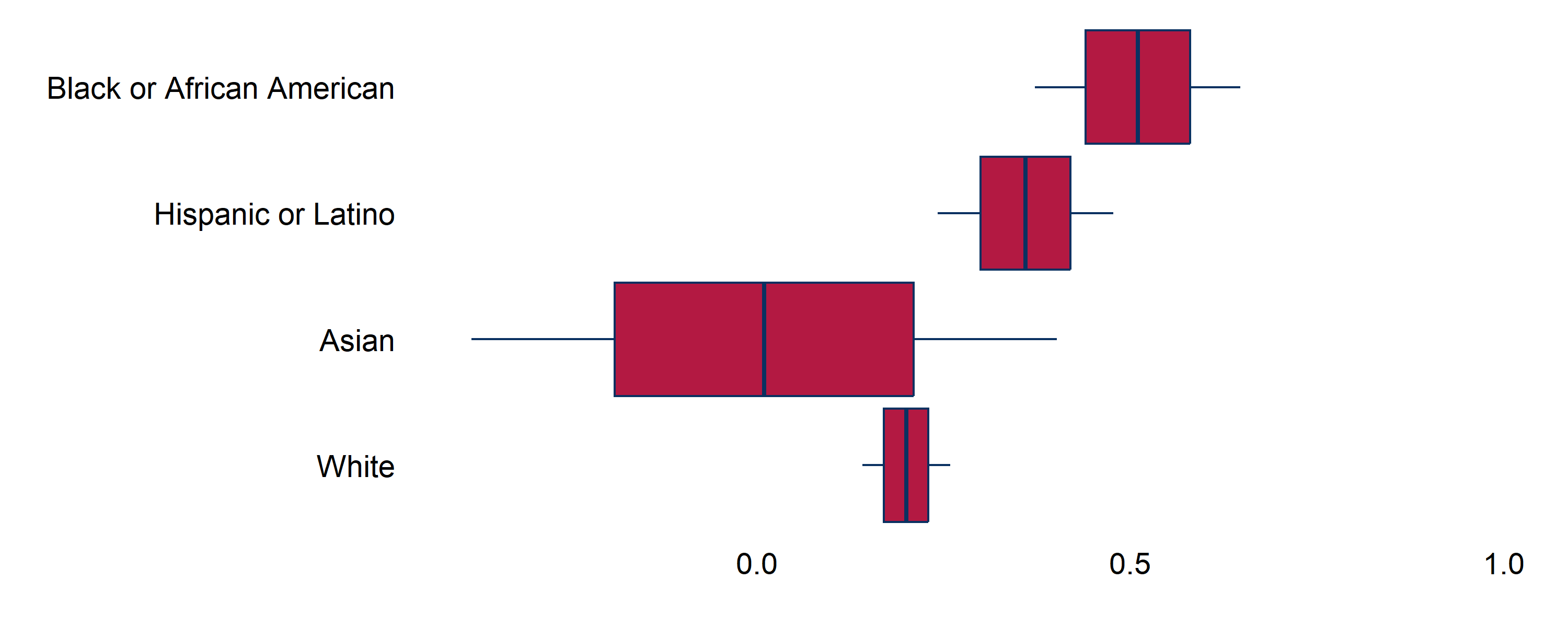}
     \end{subfigure}
     \hfill
     \begin{subfigure}[b]{0.49\textwidth}
         \centering \footnotesize
         Panel D. Overall Supply : Reason (1 year) 
         \includegraphics[width=\textwidth, height=0.5\textwidth]{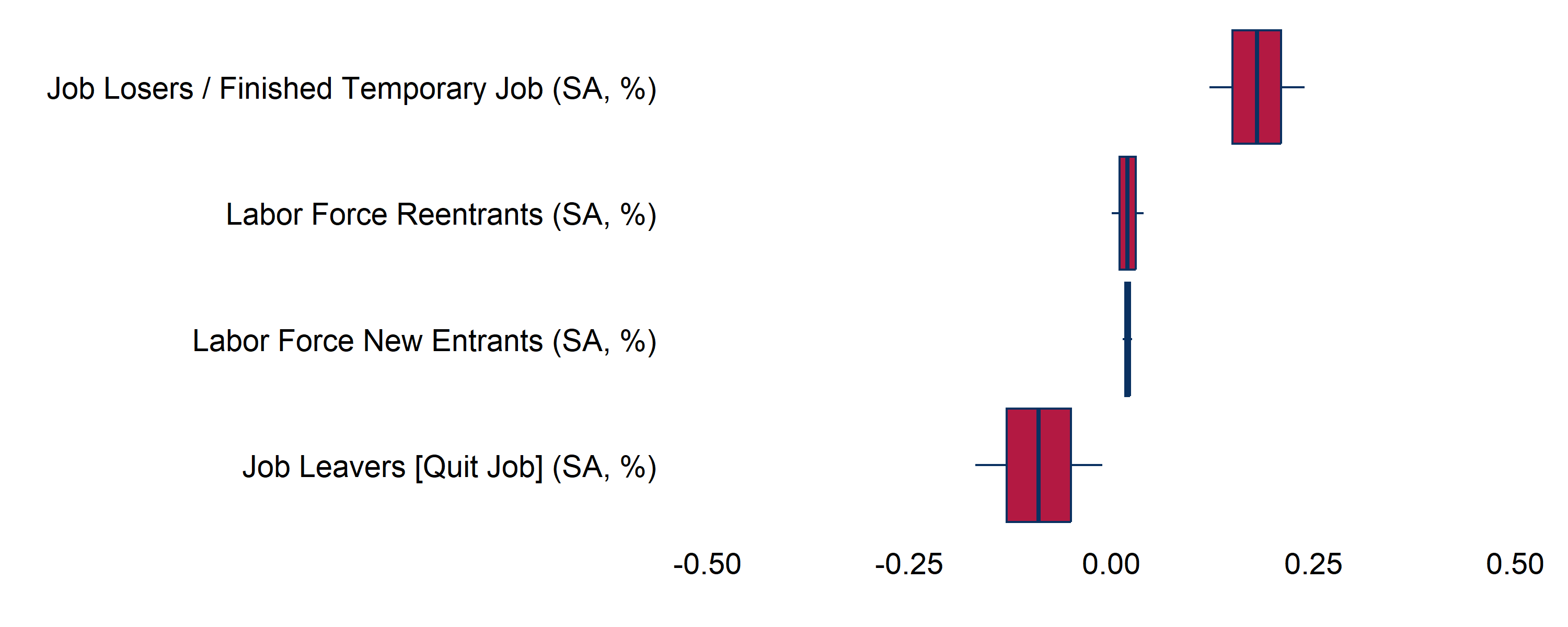}
     \end{subfigure}
 \bigskip
	\begin{subfigure}[b]{0.49\textwidth}
         \centering \footnotesize
         Panel E. Oil-Supply : Race (3 year) 
         \includegraphics[width=\textwidth, height=0.5\textwidth]{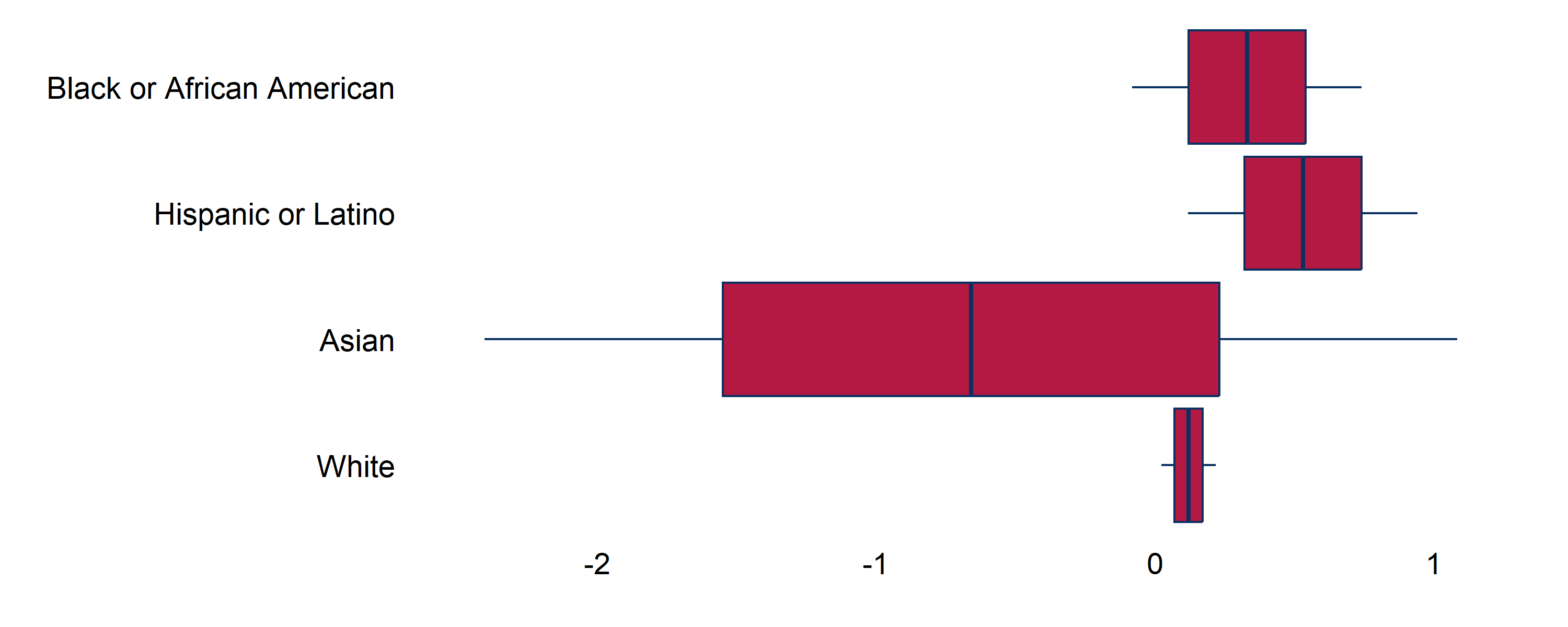}
     \end{subfigure}
     \hfill
     \begin{subfigure}[b]{0.49\textwidth}
         \centering \footnotesize
         Panel F. Oil-Supply : Reason (3 year)
         \includegraphics[width=\textwidth, height=0.5\textwidth]{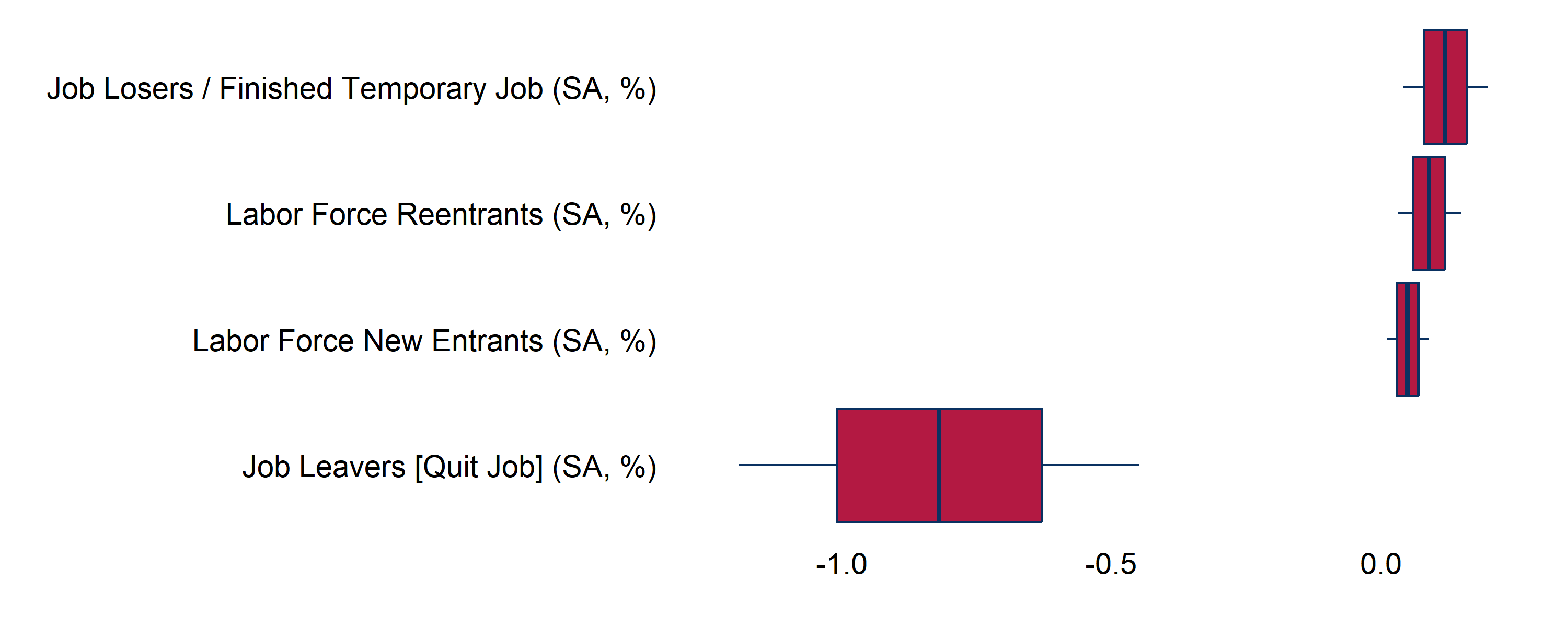}         
     \end{subfigure}
\bigskip
\begin{subfigure}[b]{0.49\textwidth}
         \centering
         Panel G. Overall Supply : Race (3 year)
         \includegraphics[width=\textwidth, height=0.5\textwidth]{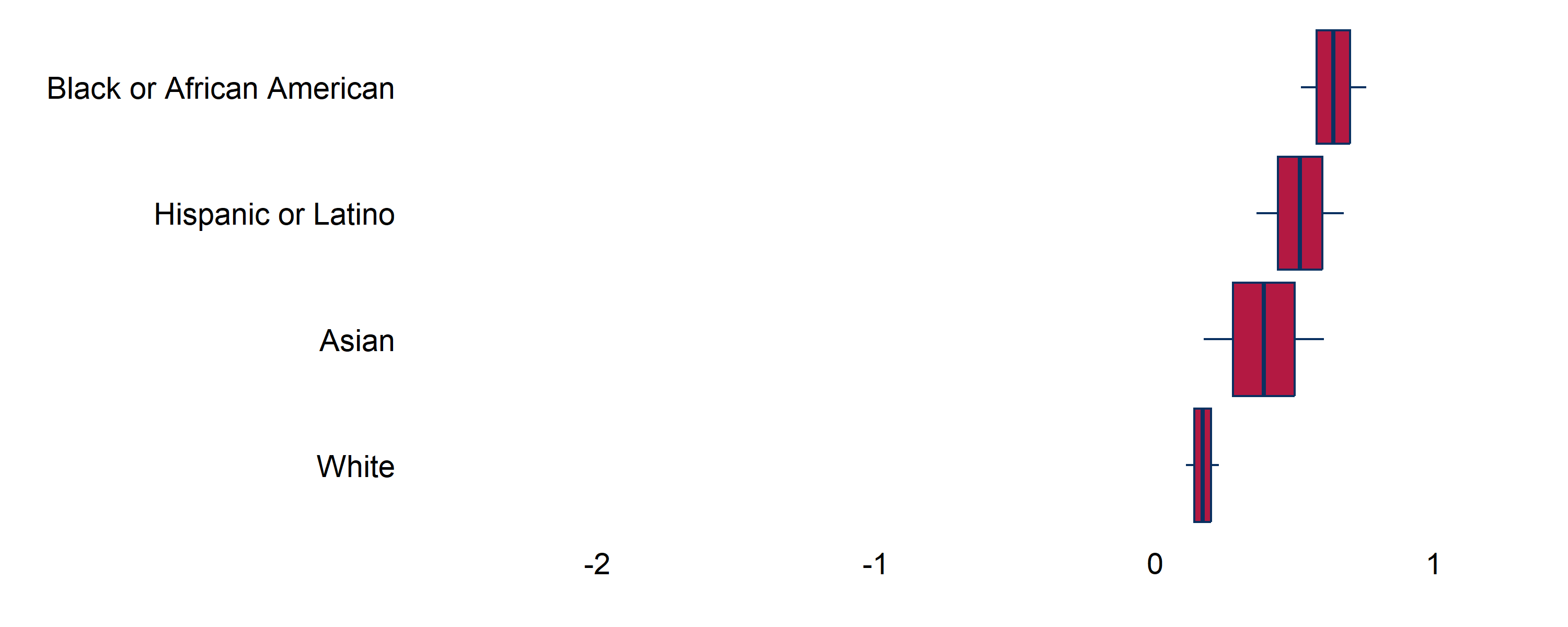}
     \end{subfigure}
     \hfill
     \begin{subfigure}[b]{0.49\textwidth}
         \centering
         Panel H. Overall Supply : Reason (3 year)
         \includegraphics[width=\textwidth, height=0.5\textwidth]{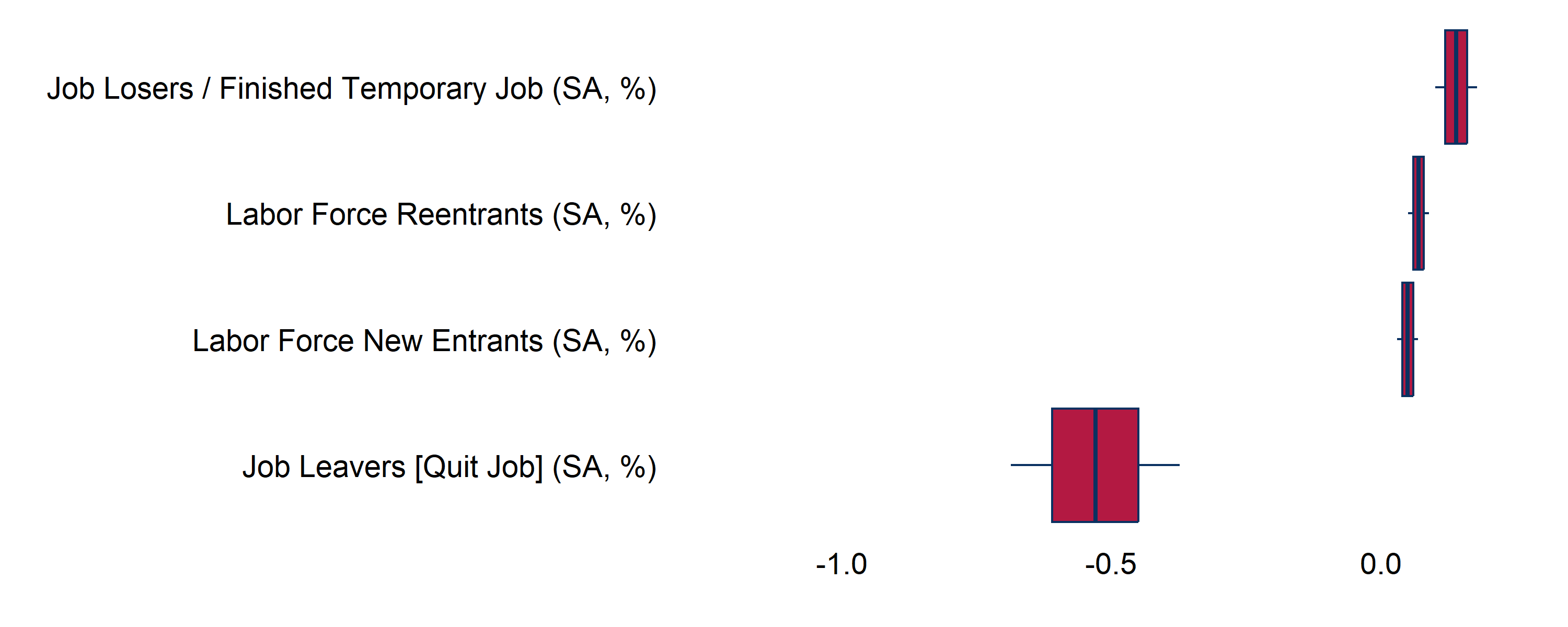}
     \end{subfigure}
\raggedright   
{\footnotesize \textbf{Notes to figure:} This table shows the response of changes in unemployment rates to a one-percentage-point increase in price inflation driven by oil supply news shock from \cite{kanzig2021macroeconomic} (Panels A, B, E, and F) and supply-driven inflation (Panels C, D, G and H). Parentheses in each panel title indicate the horizon of the unemployment rate changes. The coefficients are estimated using quantile regressions at the 80th percentile. The box plots display both one and two standard deviations. \textbf{Source:} Authors' calculation.} 
\label{figure:Oil}
\end{figure}

To evaluate the extent to which oil-supply shocks create distributional effects on unemployment tail risks, we replace the supply-driven inflation with the oil news shock by \cite{kanzig2021macroeconomic} as the instrumental variable and estimate the IVQR model. We then focus on the distributional effects by race and reason for unemployment. Panels A-D of Figure \ref{figure:Oil} display the coefficient estimates by race and reason for unemployment for one year ahead. Relative to the supply-driven inflation, the portion of inflation driven by oil supply shock generates larger and more significant differential effects along the margin of reason for unemployment between job leavers and job losers (Panels B, D, F and H), but not so much so between Blacks and Whites (Panels A, C, E and G). 

It is important to note that the portion of inflation driven by oil supply shocks has larger downside effects on the unemployment rate of job leavers than does the supply-driven inflation. This again suggests that job leavers, facing recession risks and increased pressure from rising energy prices, are reducing their job search activity. Facing the increased cost of living, even those with bargaining power who might otherwise consider quitting are less likely to voluntarily leave their jobs in search of new job opportunities. 

These results carry several implications. While supply-driven inflation is correlated with the portion of price changes attributable to oil supply shocks, the differing coefficients by race for the two types of inflation suggest that supply-side factors extend beyond just oil shocks. Second, inflation driven specifically by oil supply shocks has a more pronounced recessionary effect than broader supply-driven inflation, particularly in terms of the reason for unemployment---a category that typically shows the largest cyclical disparities in labor market outcomes. This finding reinforces that oil supply shocks are recessionary in nature and have distributional consequences akin to those of a typical recession, especially by increasing the unemployment tail risk for job leavers. 

\subsection{Monetary Policy Shock} \label{sec:MP} 


Next, we explore the relationship between demand-driven inflation and the inflationary effects of monetary policy shocks, examining the extent to which the portion of inflation driven by monetary policy influences disparities in unemployment tail risks. 

For this, we replace demand-driven inflation with the narrative-based monetary instruments from \cite{romer2004new}. The narrative-based instrument has a long historical coverage, starting from 1969 and ending in 2007.\footnote{We also consider the high-frequency monetary policy shock from \cite{bu2021unified} in Section D of the online appendix. However, the sample period for \citeauthor{bu2021unified}'s shock spans only 1994–2020, covering just three recessions. Because this period is not long enough to reliably study unemployment tail risks, we use \citeauthor{romer2004new}'s shock as the baseline.} We focus on races and reasons for unemployment. 

\begin{figure}[h!] 
	\centering 
 \caption{Effects of Monetary Policy Shock on Unemployment Risks} \footnotesize
 \medskip
	\begin{subfigure}[b]{0.49\textwidth}
         \centering \footnotesize
         Panel A. Romer-Romer : Race (1-year)
         \includegraphics[width=\textwidth, height=0.5\textwidth]{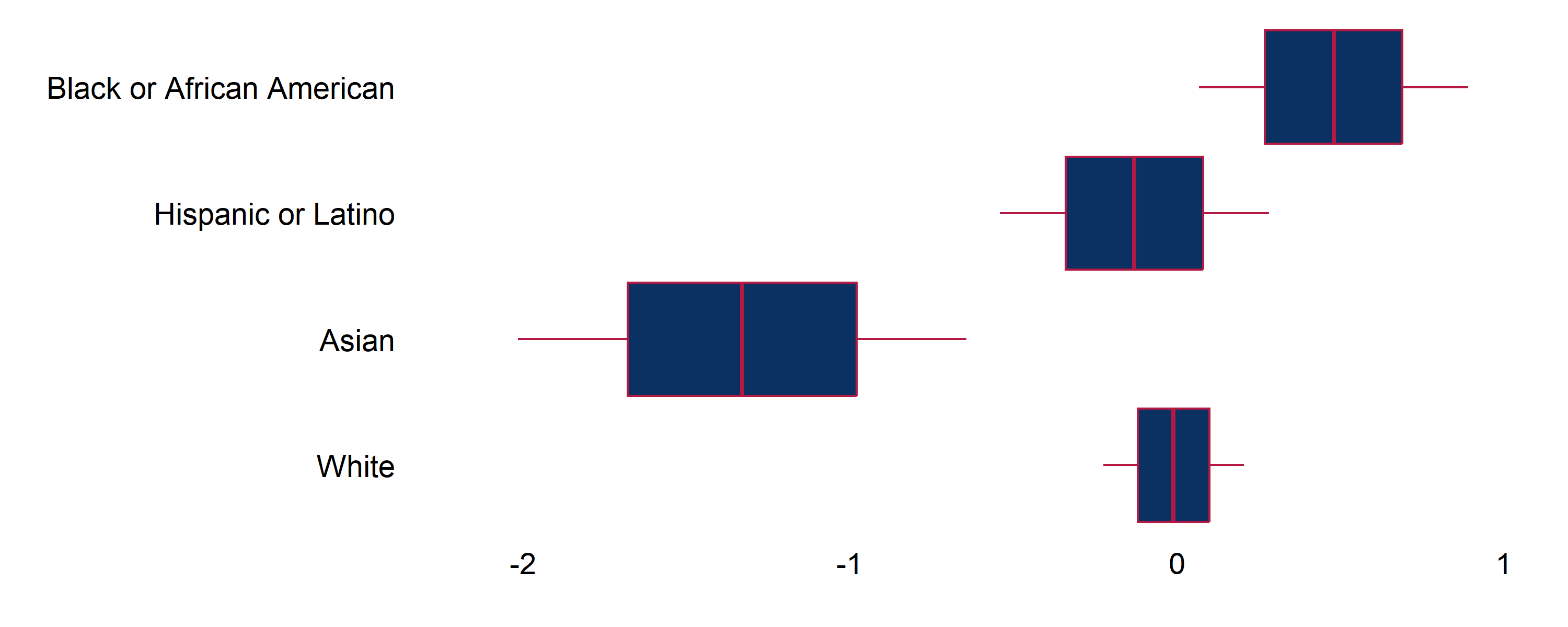}
     \end{subfigure}
     \hfill
     \begin{subfigure}[b]{0.49\textwidth}
         \centering \footnotesize
         Panel B. Romer-Romer : Reason (1-year)
         \includegraphics[width=\textwidth, height=0.5\textwidth]{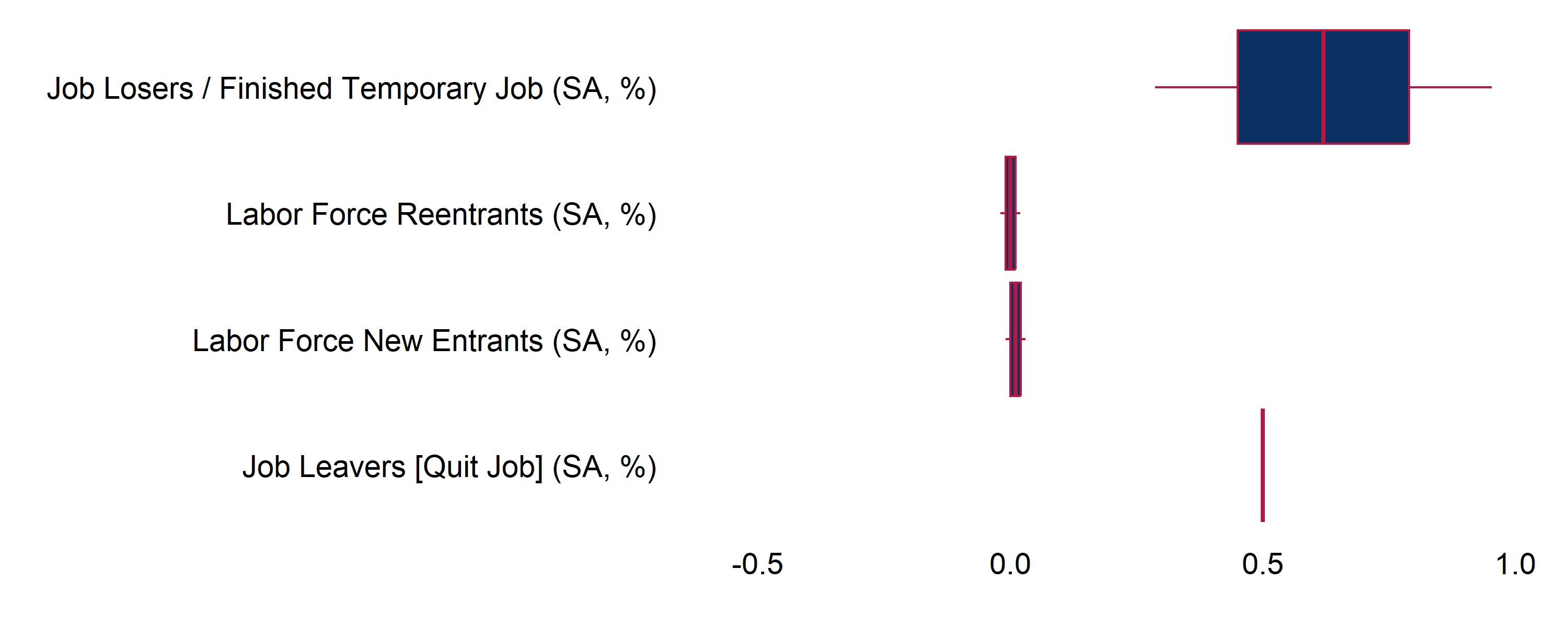} 
     \end{subfigure}     
\bigskip
\begin{subfigure}[b]{0.49\textwidth}
         \centering \footnotesize
         Panel C. Overall Demand : Race (1-year)
         \includegraphics[width=\textwidth, height=0.5\textwidth]{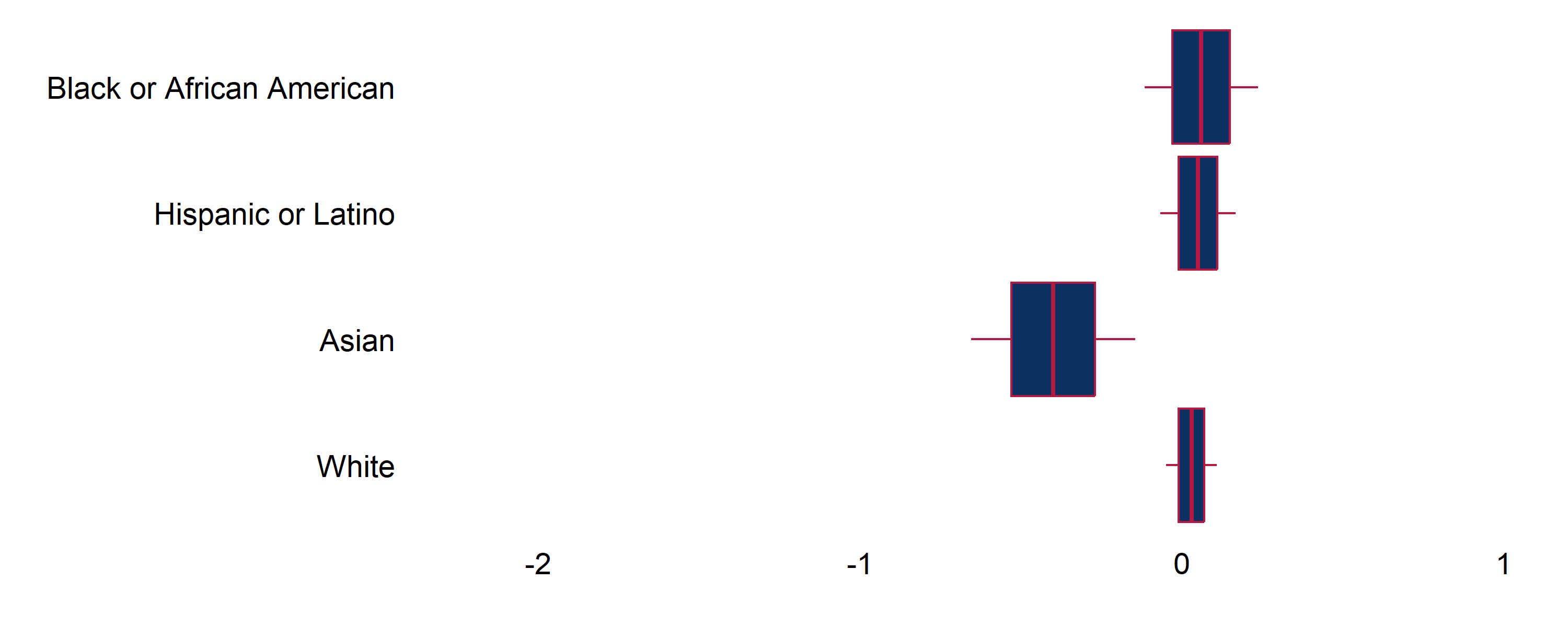}
     \end{subfigure}
     \hfill
     \begin{subfigure}[b]{0.49\textwidth}
         \centering \footnotesize
         Panel D. Overall Demand : Reason (1-year)
         \includegraphics[width=\textwidth, height=0.5\textwidth]{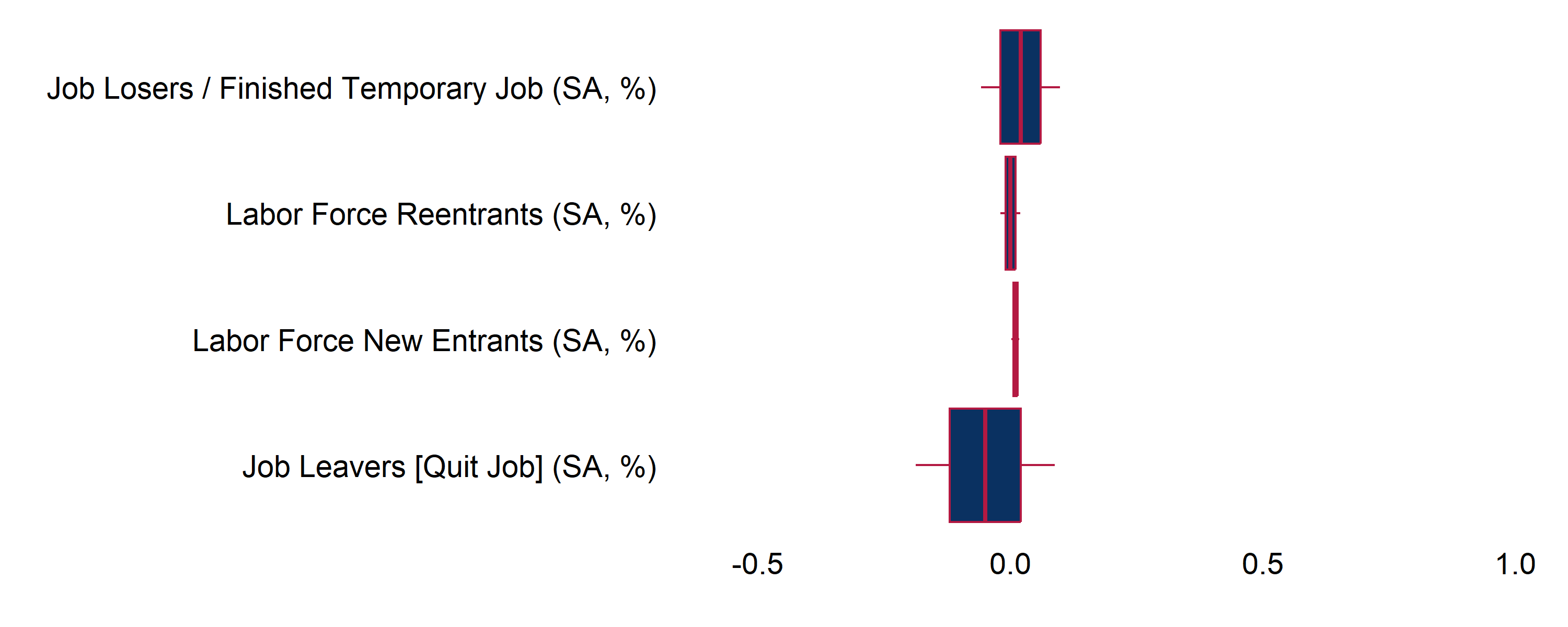}
     \end{subfigure}    
	\begin{subfigure}[b]{0.49\textwidth}
         \centering \footnotesize
         Panel E. Romer-Romer : Race  (3-year)
         \includegraphics[width=\textwidth, height=0.5\textwidth]{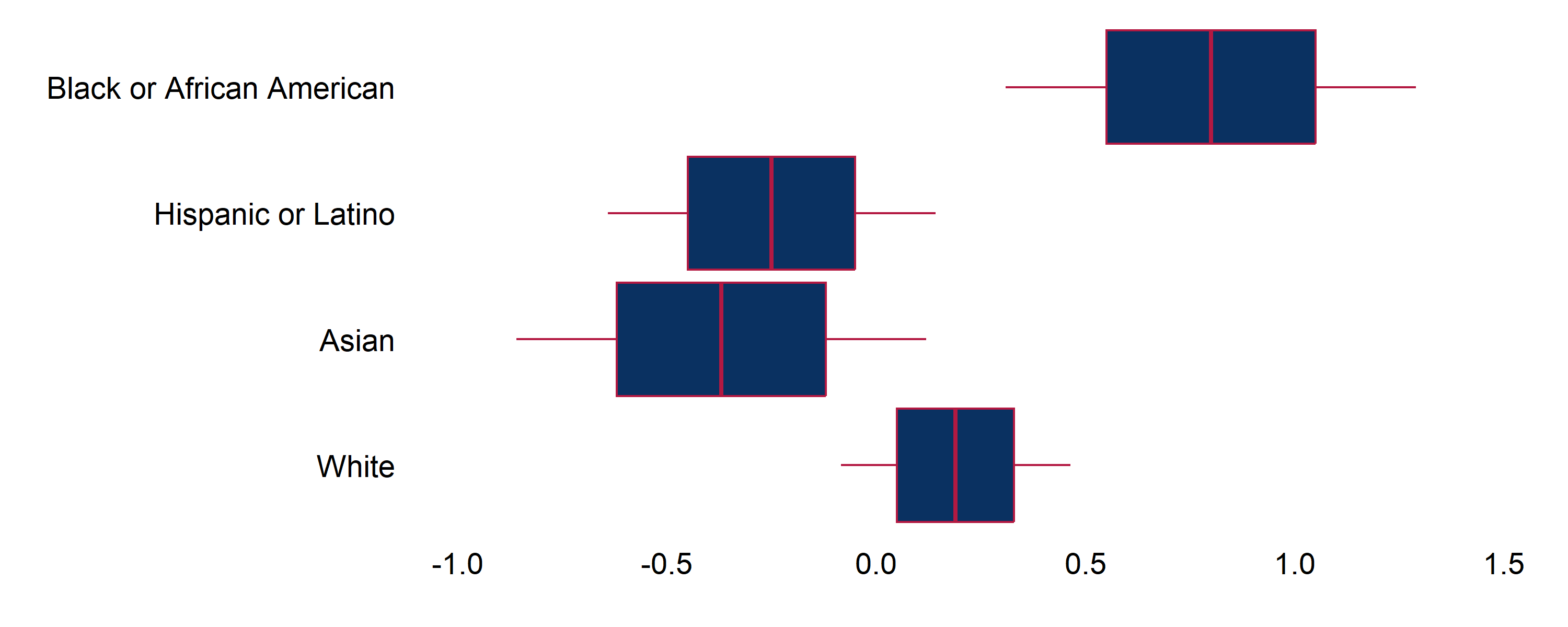}
     \end{subfigure}
     \hfill
     \begin{subfigure}[b]{0.49\textwidth}
         \centering \footnotesize
         Panel F. Romer-Romer : Reason (3-year)
         \includegraphics[width=\textwidth, height=0.5\textwidth]{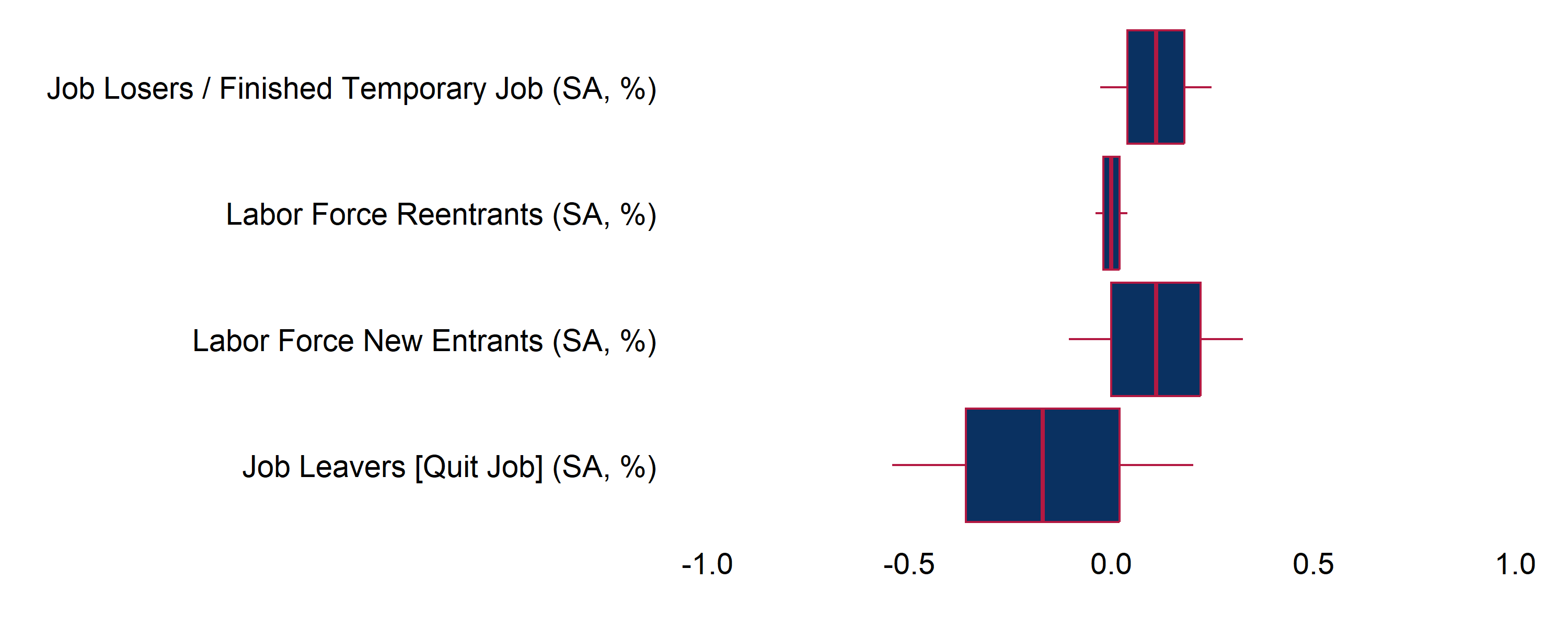}     
     \end{subfigure}     
\bigskip
\begin{subfigure}[b]{0.49\textwidth}
         \centering \footnotesize
         Panel G. Overall Demand : Race  (3-year)
         \includegraphics[width=\textwidth, height=0.5\textwidth]{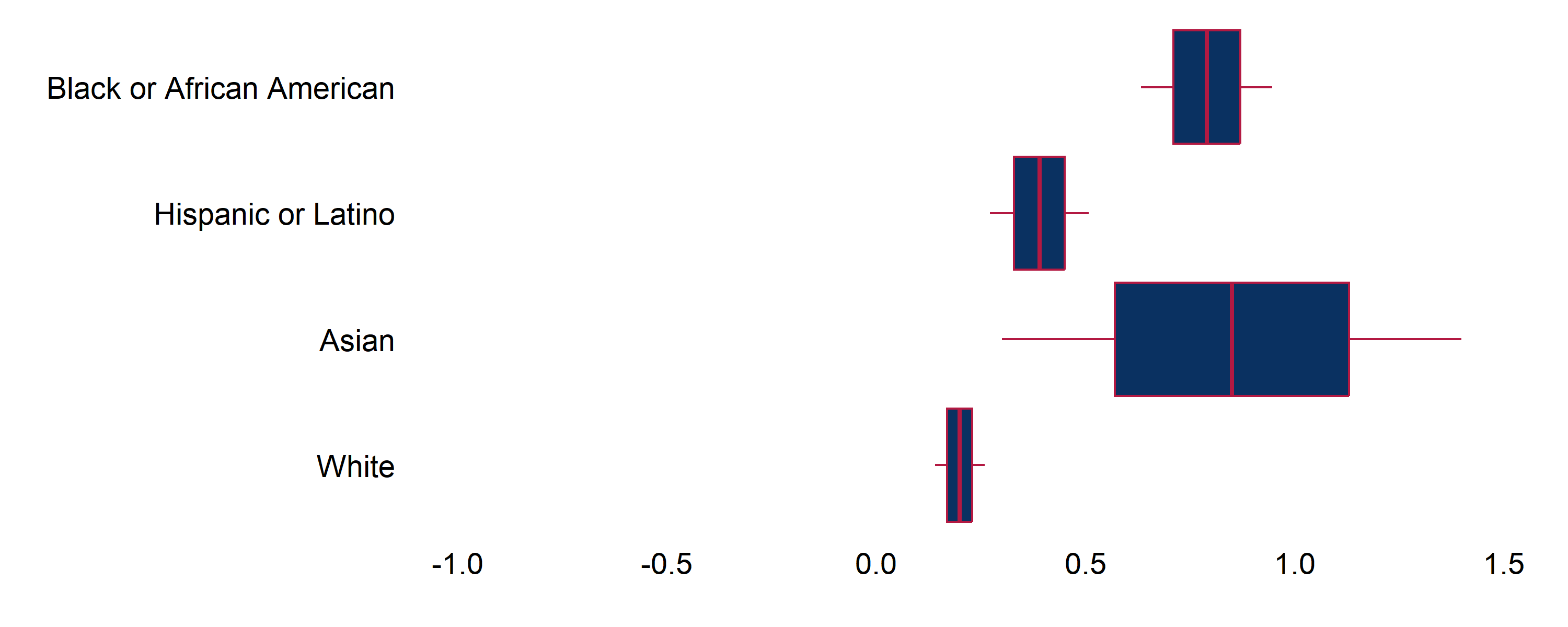}
     \end{subfigure}
     \hfill
     \begin{subfigure}[b]{0.49\textwidth}
         \centering \footnotesize
         Panel H. Overall Demand : Reason (3-year)
         \includegraphics[width=\textwidth, height=0.5\textwidth]{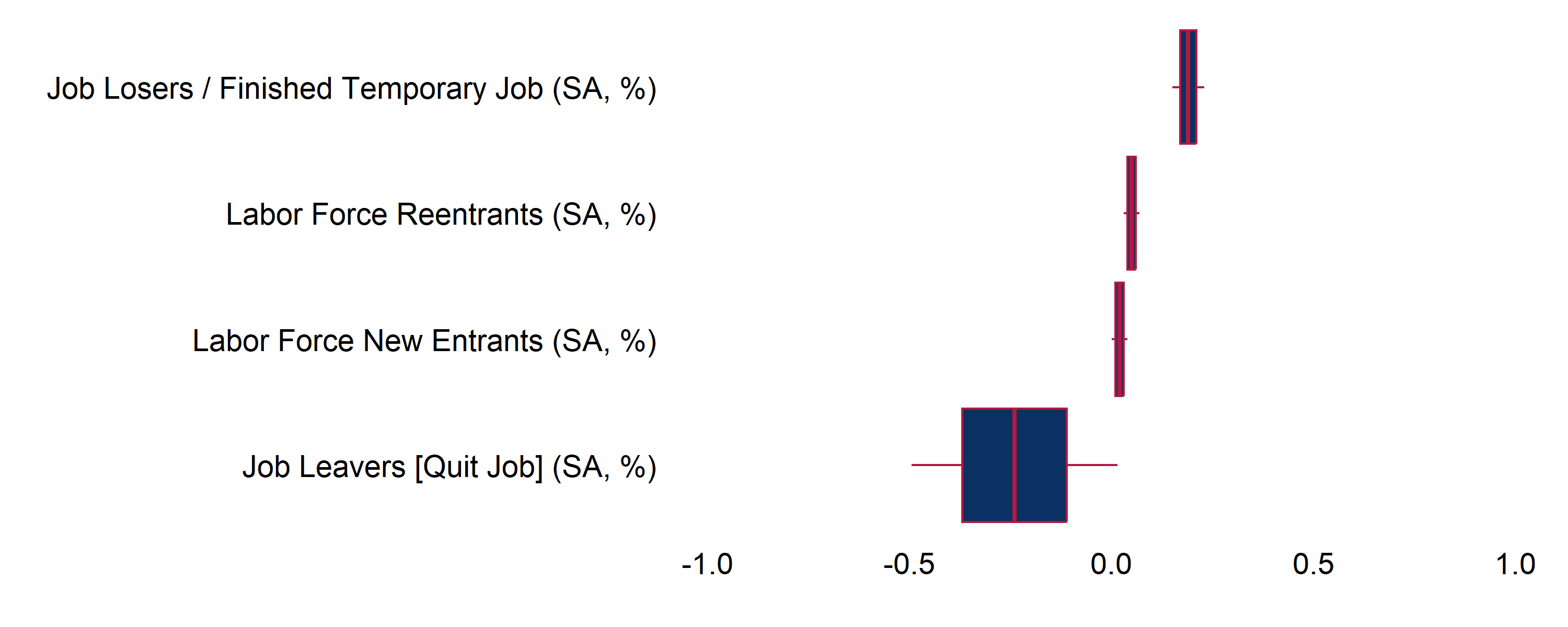}
     \end{subfigure}  

\raggedright 
{\footnotesize \textbf{Notes to figure:} This table shows the response of changes in unemployment rates to a one-percentage-point increase in price inflation driven by expansionary monetary policy shock from \cite{Romer_Romer1989} (Panels A, B, E, and F) and demand-driven inflation (Panels C, D, G and H). Parentheses in each panel title indicate the horizon of the unemployment rate changes. The coefficients are estimated using quantile regressions at the 80th percentile. The box plots display both one and two standard deviations. \textbf{Source:} Authors' calculation.} \label{figure:RR}
\end{figure}

Figure \ref{figure:RR} reports the parameter estimates with two externally identified monetary policy shocks (Panels A, B, E, and F) along with the results for demand-driven inflation (Panels C, D, G, and H). Notably, the portion of inflation driven by expansionary monetary policy has statistically significant distributional effects in both the short and medium run. Quite differently, the overall demand-driven inflation has effects only in the medium run. 
In particular, the differential effects are apparent across racial groups and workers unemployed for different reasons for unemployment. Inflation driven by monetary policy shock raises the unemployment tail risk of Blacks more than that of Whites, and that of job losers more than that of entrants to the labor force.\footnote{When instrumenting with the shock of \citeauthor{bu2021unified} (henceforth, BRW), the adverse inflationary effects on job losers' unemployment tail risks relative to others' remain statistically significant despite the short sample period of the shock.}

All told, the empirical results have important implications for monetary policy. Inflation driven by expansionary monetary policy raises the unemployment risks of racial minorities and job losers both in the short term and the medium term. The conduct of expansionary monetary policy to reduce the employment shortfalls of racial minorities and cyclically vulnerable workers can instead raise their unemployment tail risks through the inflation channel, unintentionally exacerbating inequality in the labor market. This result indicates a risk of running the economy hot for the equitable growth and highlights that stabilizing inflation also reduces the unemployment tail risk of racial minority. 

\subsection{Wage Inflation} \label{sec:wage} 

The observed distributional effects of both supply- and demand-driven inflation on unemployment tail risks highlight wage setting and wage inflation expectations as potentially important underlying factors. Wage inflation itself can be both a supply- and demand-side driver of price inflation. An increase in the minimum wage can directly push up the prices of services, while wage inflation driven by stronger labor demand can be passed through to overall price inflation.\footnote{\cite{kiley2023role} find that wage inflation is an important predictor of price inflation during high-inflation periods. In addition, \cite{Gali2011} argues that the wage Phillips curve is steeper than its price counterpart, implying that wage inflation may capture labor demand shocks well even without conditioning on them.} To examine the role of wage inflation, we replace price inflation with nominal wage inflation in a reduced-form quantile regression without using instrumental variables. As our wage measure, we use average hourly earnings of production and non-supervisory workers, available from January 1964 to the most recent month.\footnote{In Section D of the online appendix, we use the Atlanta Wage Tracker, which provides wage growth data by worker characteristics, to distinguish the effect of each group’s wage inflation on its respective unemployment tail risk. The results confirm the robustness of our main findings. In addition, we also consider labor supply and demand shocks from \cite{baumeister2015sign} as instruments, as discussed in Section E of the online appendix. The results show that these shocks differ from the supply and demand shocks that raise price inflation.}

\begin{figure}
	\centering
 \caption{Unemployment Risk: Wage Inflation} 
 \medskip
 \footnotesize
 \begin{subfigure}[b]{0.49\textwidth}
         \centering \footnotesize
         Panel A. Race -- wage inflation (1-year)
         \includegraphics[width=\textwidth, height=0.5\textwidth]{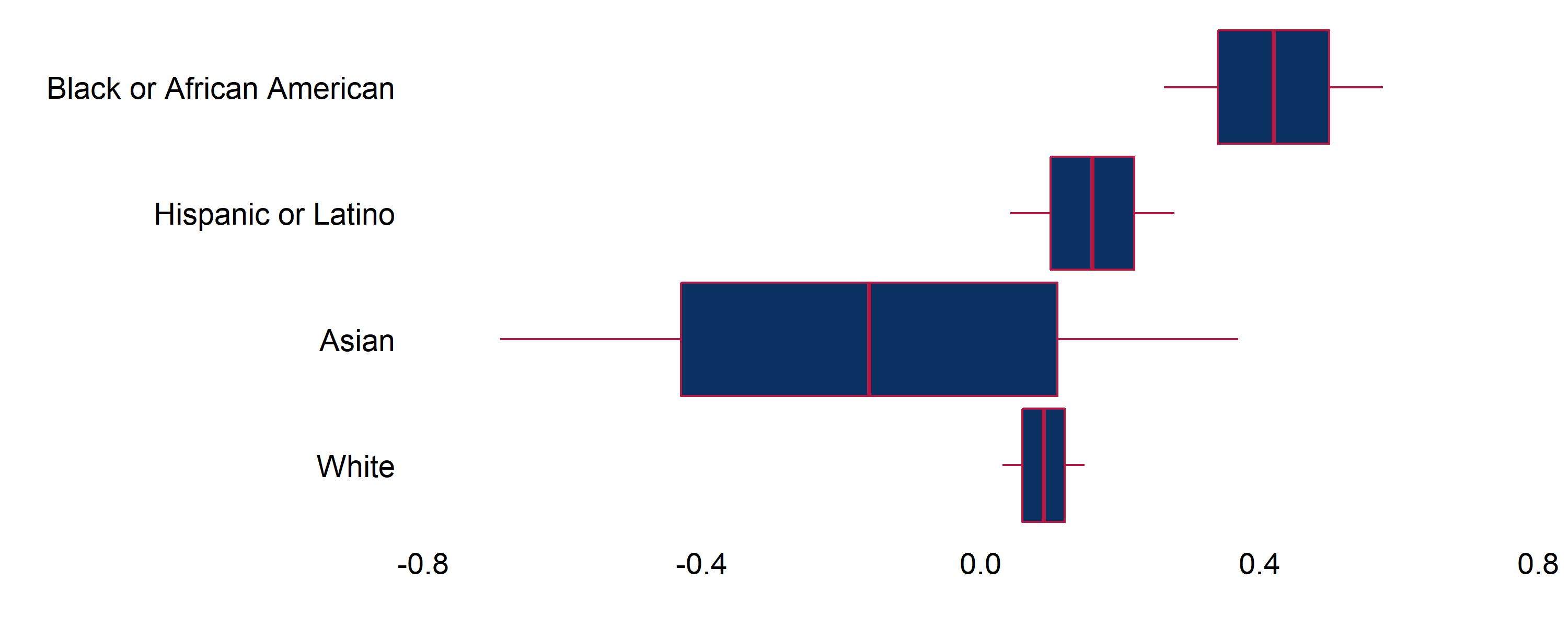}
     \end{subfigure}
     \hfill
       \begin{subfigure}[b]{0.49\textwidth}
         \centering \footnotesize
         Panel B. Race -- wage inflation (3-year)
         \includegraphics[width=\textwidth, height=0.5\textwidth]{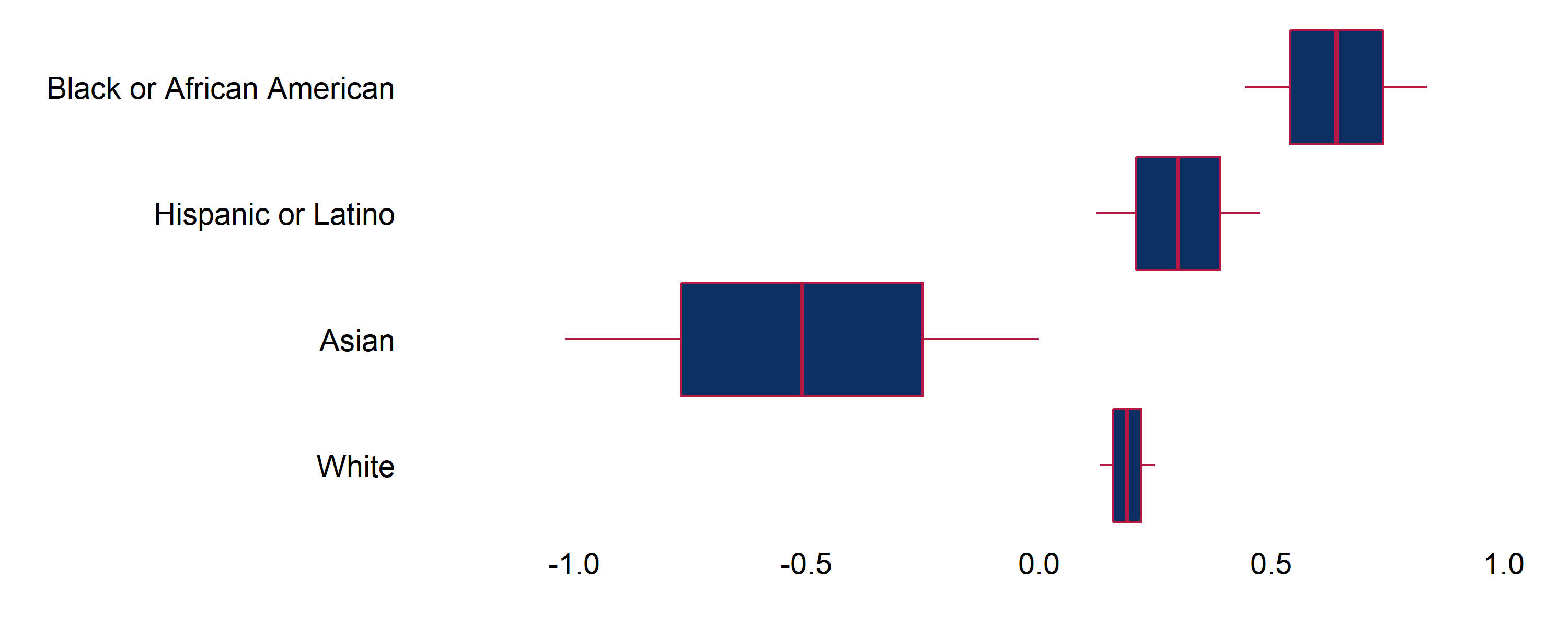}
     \end{subfigure}
\hfill 
\begin{subfigure}[b]{0.49\textwidth}
         \centering
         {\footnotesize C. Race-Supply (1 year)} 
         \includegraphics[width=1.5\textwidth, height=0.49\textwidth]{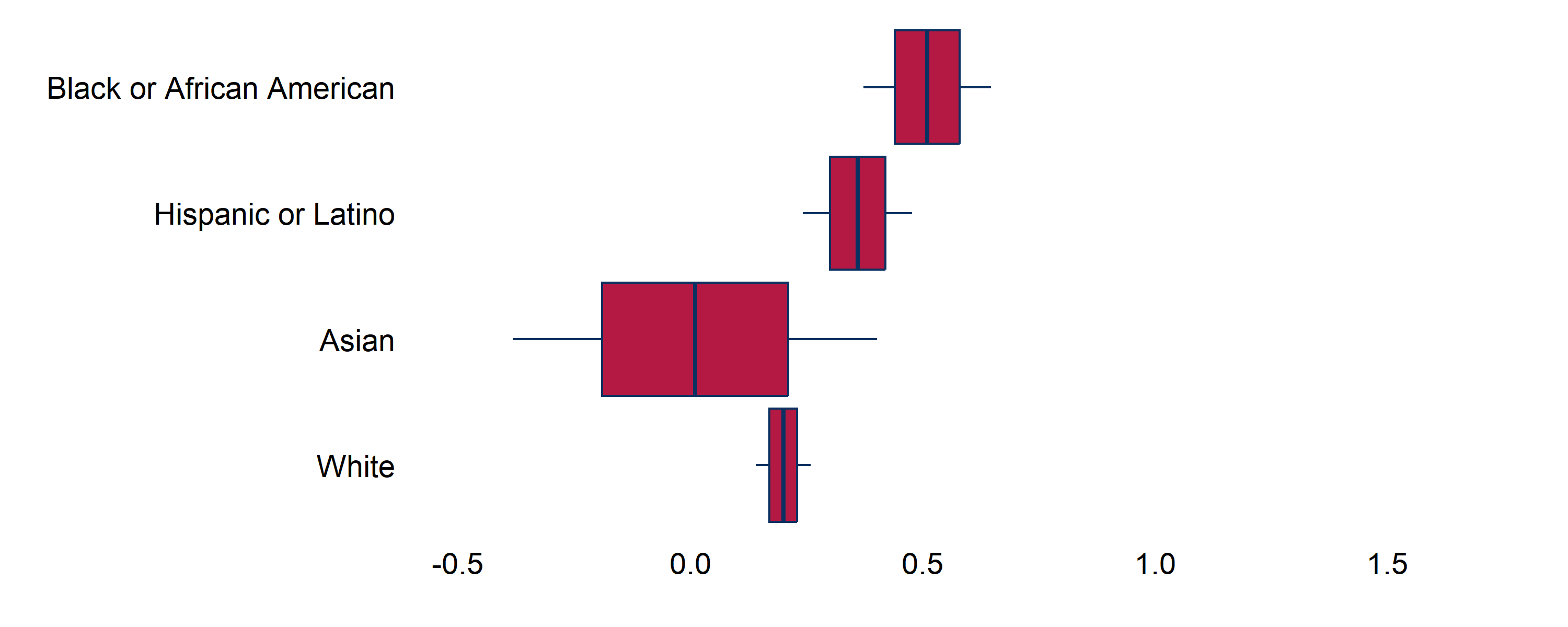}
     \end{subfigure}
     \hfill
\begin{subfigure}[b]{0.49\textwidth}
         \centering
         {\footnotesize D. Race-Demand (3 year)} 
         \includegraphics[width=1.15\textwidth, height=0.49\textwidth]{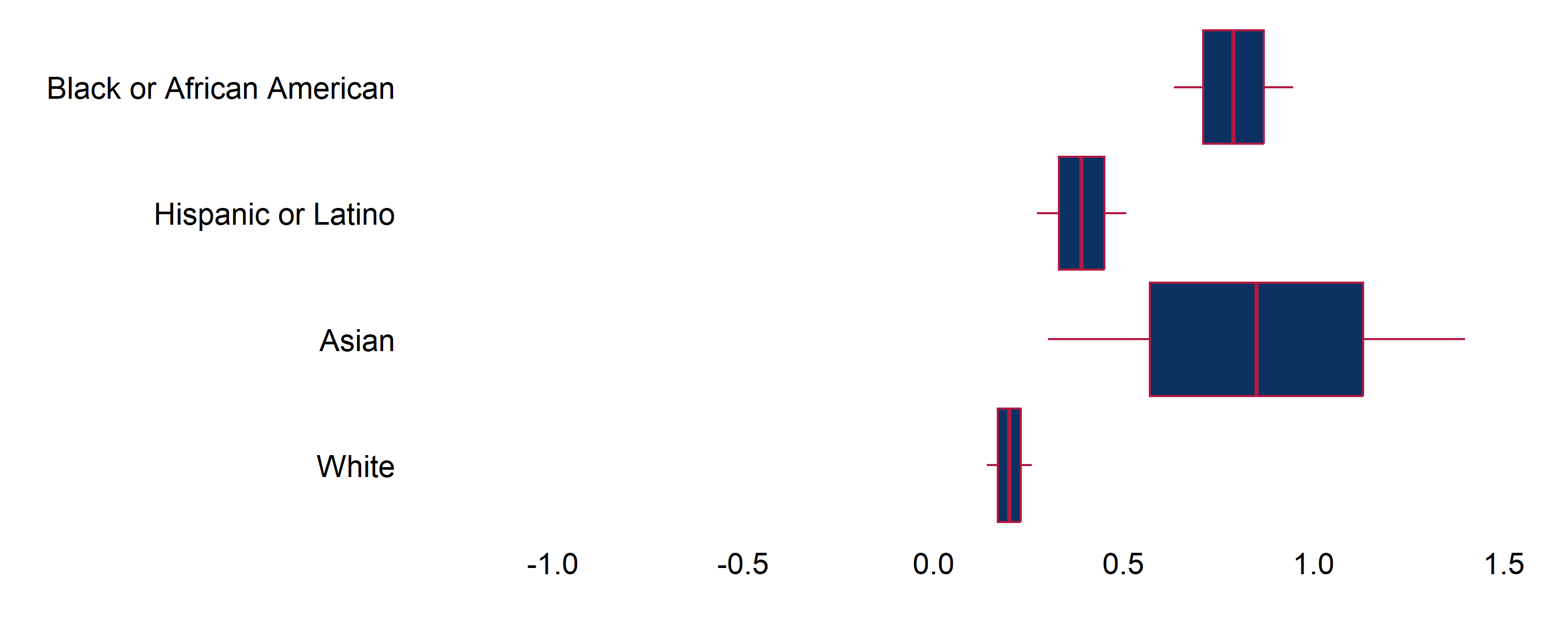}
\end{subfigure}

\begin{subfigure}[b]{0.49\textwidth}
         \centering \footnotesize
         Panel E. Reasons -- wage inflation (1-year)
         \includegraphics[width=\textwidth, height=0.5\textwidth]{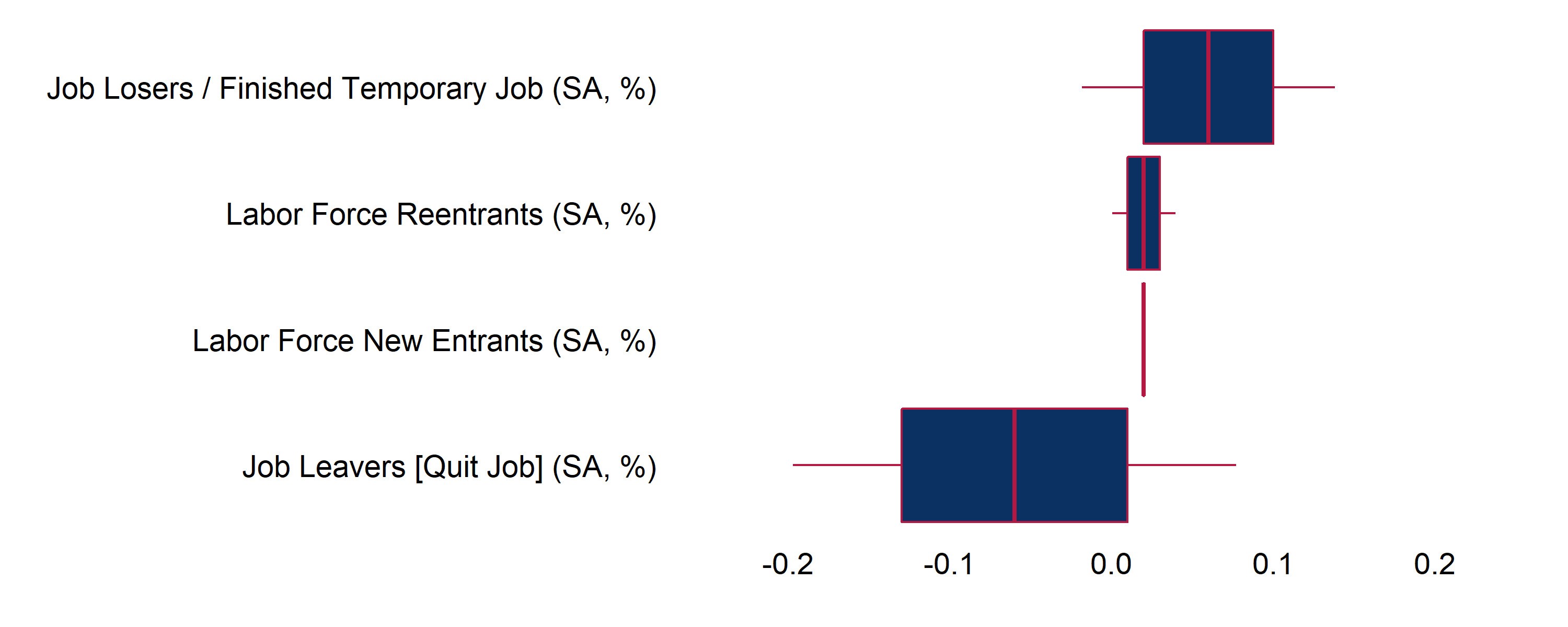}
     \end{subfigure}
     \hfill    
     \begin{subfigure}[b]{0.49\textwidth}
         \centering \footnotesize
         Panel F. Reasons -- wage inflation (3-year)
         \includegraphics[width=\textwidth, height=0.5\textwidth]{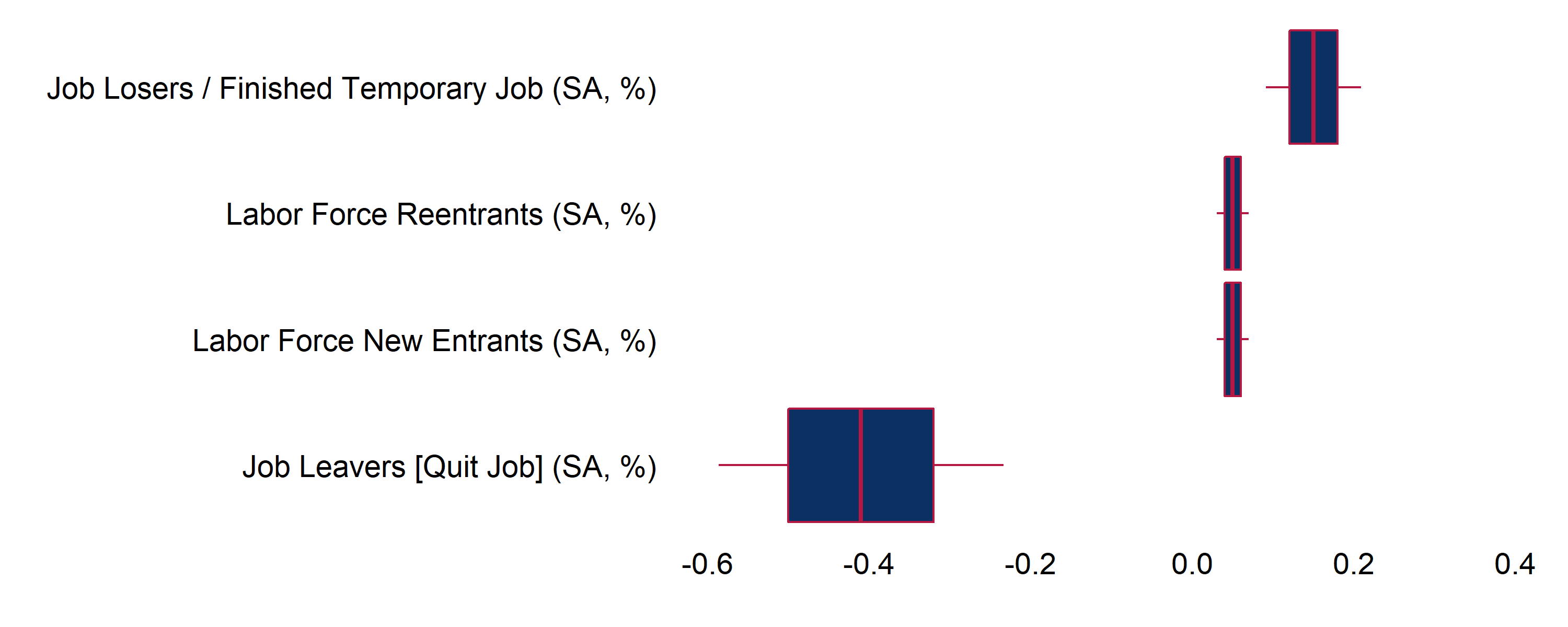}
     \end{subfigure}

     \begin{subfigure}[b]{0.49\textwidth}
         \centering
         {\footnotesize G. Reason-Supply (1 year)} 
         \includegraphics[width=1.05\textwidth, height=0.49\textwidth]{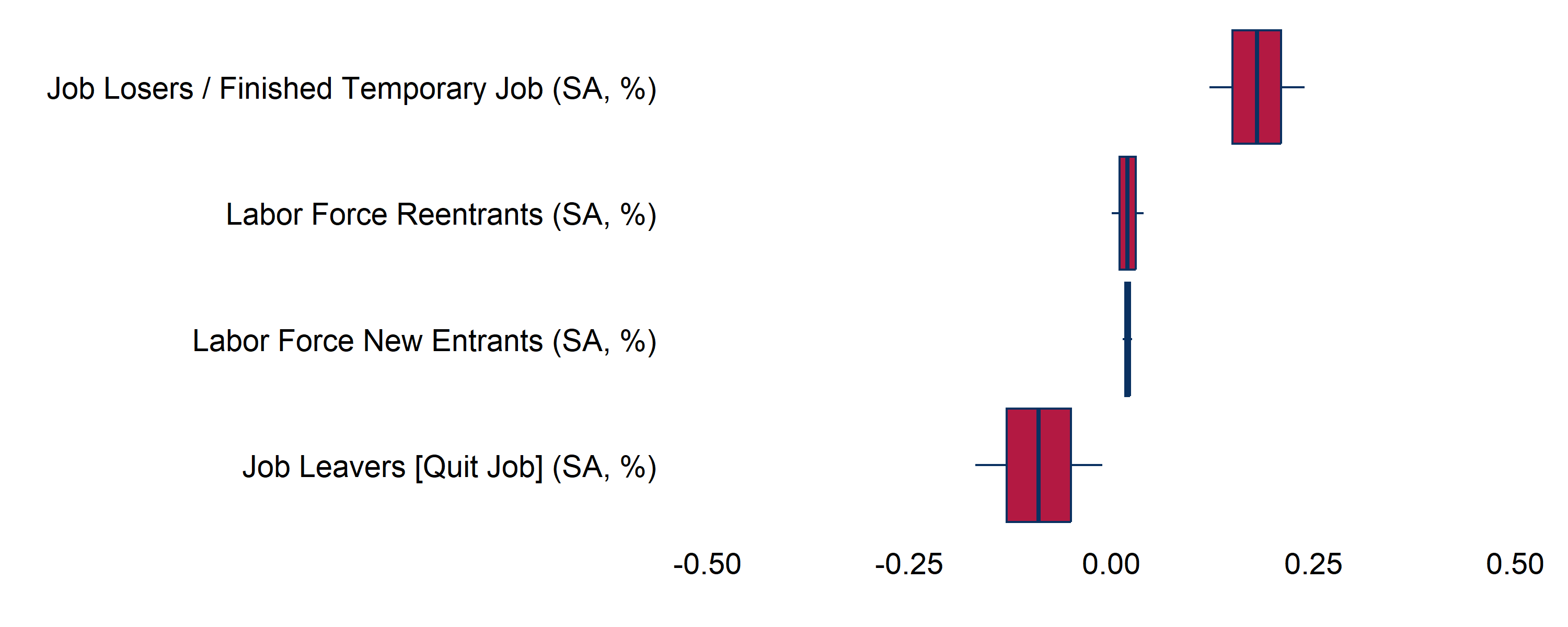}
     \end{subfigure}
     \hfill
\begin{subfigure}[b]{0.49\textwidth}
         \centering
         {\footnotesize H. Reason-Demand (3 year)} 
         \includegraphics[width=1.13\textwidth, height=0.49\textwidth]{Q80IVQR_HINFD_3year_Reasonsfig14.png}
\end{subfigure}
\raggedright 
{\footnotesize \textbf{Notes to figure:} This table shows the response of changes in unemployment rates to a one-percentage-point increase in wage inflation (Panels A, B, E, and F), supply-driven inflation (Panels C and G), and demand-driven inflation (Panels D and H). Parentheses in each panel title indicate the horizon of the unemployment rate changes. The coefficients are estimated using quantile regressions at the 80th percentile. The box plots display both one and two standard deviations. \textbf{Source:} Authors' calculation.} \label{figure:Q80QR_WINF_3year}
\end{figure}

Figure \ref{figure:Q80QR_WINF_3year} presents the estimation results by race (Panels A–D) and by reason for unemployment (Panels E–H). Notably, the statistically significant differential effects of wage and supply-driven price inflation are similar across racial groups one year ahead.\footnote{The sample period for Asians is shorter than that of other groups, resulting in greater uncertainty in the estimates, as reflected in their wider confidence intervals.} Specifically, a one percentage point increase in wage inflation raises the unemployment rate of Black individuals by about half a percentage point, with much smaller effects on the unemployment rate of White individuals---similar in magnitude to the effects of supply-driven inflation (approximately 0.1–0.2 percentage points). For three years ahead, both wage inflation and demand-driven inflation exhibit similar distributional effects across racial groups and by reason for unemployment. A one percentage point rise in wage inflation increases the unemployment rate of Black individuals by 0.6 percentage point and that of White individuals by 0.2 percentage point, broadly in line with the effects of demand-driven inflation. Furthermore, a one percentage point increase in wage inflation raises the unemployment rate of job losers by 0.2 percentage points but lowers the unemployment rate of job leavers by 0.4 percentage points. Demand-driven inflation shows similar effects for job losers, while its impact on job leavers’ unemployment is slightly smaller, reducing it by about a quarter percentage point. 

In summary, wage inflation resembles supply-driven inflation in the short term, but mirrors demand-driven inflation in the medium term in its cross-sectional effects by race and reason for unemployment. These results suggest that wage inflation plays a significant role in the distributional effects of price inflation on unemployment tail risk, operating through both supply and demand channels.

What explains the observed tight link between wage inflation and price inflation of the two types? To begin with, as mentioned in the opening discussion of this section, wage inflation itself can act as a cost-push factor (e.g., through changes in minimum wages) and simultaneously reflect the demand factor of inflation, as suggested by the relatively steep wage Phillips curve (\citealp{Gali_Gambetti2019}). And thus, wage inflation itself has features of both supply and demand inflation. Next, both supply- and demand-side sources of inflation can raise wage inflation expectations. Workers at the margins of the labor force---including racial and ethnic minorities---are more likely to hold higher wage inflation expectations than others, as they tend to be more exposed to higher and more volatile price changes. For example, \cite{Lee2022} finds that racial and ethnic minorities face higher inflation rates than Whites. \cite{Orchard2022} shows that low-income workers are exposed more to prices of necessities that are largely determined by energy and food prices.

Higher labor costs raise the likelihood of layoffs. The effects of supply-driven inflation tend to materialize more immediately in the short term, as firms facing rising input costs often lay off workers with low productivity or skills. In addition, these workers are likely to demand higher wages since they are exposed more to necessity prices largely determined by food and energy prices (\citealp{Orchard2022}), which raises their unemployment risks further. In contrast, the effects of demand-driven inflation unfold more gradually. Strong demand persistently raises expected wage inflation. While firms may initially retain existing workers to meet increased demand despite rising wage inflation, sustained upward pressure on labor costs can incentivize the adoption of labor-saving technologies, which eventually displace workers with low productivity or skills. All told, heterogeneity in experienced inflation and wage inflation expectations can be an important channel through which inflation generates distributional effects on unemployment tail risks.


\section{Conclusion}\label{sec:conclusion}

This paper explores the distributional effects of inflation on workers’ unemployment tail risks using quantile regression with instrumental variables. By differentiating between supply-driven and demand-driven inflation, we find that supply-driven inflation has more pervasive and immediate distributional effects. In the medium term, supply-driven inflation affects all categories considered, disproportionately raising unemployment tail risks for cyclically vulnerable groups. Meanwhile, the distributional effects of demand-driven inflation are relatively muted in the short run but become more pronounced in the medium term. Among the various worker attributes considered---including demographic characteristics, education, job status, and reason for unemployment---cross-sectional heterogeneity is most pronounced across racial groups and reasons for unemployment.

These empirical results have important implications for demand policies such as monetary policy and fiscal policy. Policies aimed at boosting demand, including monetary policy, may inadvertently exacerbate unemployment tail risks for racial minorities and job losers through the inflation channel. Consequently, while demand-boosting policies intended to reduce employment shortfalls among racial minorities and cyclically vulnerable workers can rather worsen labor market inequality because of their inflationary nature. In this context, maintaining inflation stability is essential when implementing policies aimed at promoting equitable growth in the labor market.

Lastly, wage inflation resembles supply-driven inflation in the short term but mirrors demand-driven inflation in the medium term in its cross-sectional effects by race and reason for unemployment. These results suggest that wage inflation plays a significant role in the distributional effects of price inflation on unemployment tail risk, operating through both supply and demand channels. We also discuss how heterogeneity in experienced inflation and wage inflation expectations may serve as important channels through which inflation generates distributional effects on unemployment tail risks. Further empirical and theoretical investigation into these structural mechanisms could be a valuable direction for future research.
 
\clearpage
\onehalfspacing 

{{
\setlength{\bibsep}{.2cm}
\bibliographystyle{apalike}
\bibliography{reference}
}}
\end{document}